\documentclass{jfm}
\usepackage{graphicx}
\usepackage{epstopdf, epsfig}

\usepackage[dvipsnames]{xcolor}
\usepackage{color,soul}
\usepackage{siunitx}

\shorttitle{Unified scaling for segregation forces}
\shortauthor{L. Jing, J. M. Ottino, R. M. Lueptow, and P. B. Umbanhowar}

\title{A unified description of gravity- and kinematics-induced segregation forces in dense granular flows}

\author{Lu  Jing\aff{1},
  Julio M.  Ottino\aff{1,2,3},
  Richard M.  Lueptow\aff{1,2,3}  
   \corresp{\email{r-lueptow@northwestern.edu}},\\
\and Paul B.  Umbanhowar\aff{2}}

\affiliation{
\aff{1}Department of Chemical and Biological Engineering, Northwestern University, Evanston, IL 60208, USA
\aff{2}Department of Mechanical Engineering, Northwestern University, Evanston, IL 60208, USA
\aff{3}Northwestern Institute on Complex Systems (NICO), Northwestern University, Evanston, IL 60208, USA
}

\begin{document}

\maketitle

\begin{abstract}
Particle segregation is common in natural and industrial processes involving flowing granular materials. Complex, and seemingly contradictory, segregation phenomena have been observed for different boundary conditions and forcing. Using discrete element method simulations, we show that segregation of a single particle intruder can be described in a unified manner across different flow configurations. A scaling relation for the net segregation force is obtained by measuring forces on an intruder particle in controlled-velocity flows where gravity and flow kinematics are varied independently. The scaling law consists of two additive terms: a buoyancy-like gravity-induced pressure gradient term and a shear rate gradient term, both of which depend on the particle size ratio. The shear rate gradient term reflects a kinematics-driven mechanism whereby larger (smaller) intruders are pushed toward higher (lower) shear rate regions. The scaling is validated, without refitting, in wall-driven flows, inclined wall-driven flows, vertical silo flows, and free surface flows down inclines. Comparing the segregation force to the intruder weight results in predictions of the segregation direction that match experimental and computational results for various flow configurations.
\end{abstract}

% \begin{keywords}
% % Authors should not enter keywords on the manuscript, as these must be chosen by the author during the online submission process and will then be added during the typesetting process (see http://journals.cambridge.org/data/\linebreak[3]relatedlink/jfm-\linebreak[3]keywords.pdf for the full list)
% \end{keywords}

\section{Introduction}

Flowing granular mixtures tend to segregate by size, density, or other physical properties. Understanding granular segregation is essential for industrial sectors where granular materials are mixed and de-mixed \citep{ottino_mixing_2000,ottino_mixing_2008}. Segregation also plays an important role in natural processes such as geophysical mass flows \citep{iverson_physics_1997,johnson_grain-size_2012} and bedload transport \citep{frey_how_2009,ferdowsi_river-bed_2017}.
% geophysical mass flows \citep{iverson_physics_1997,felix_relation_2004,johnson_grain-size_2012,kokelaar_fine-grained_2014} and bedload transport \citep{frey_how_2009,ferdowsi_river-bed_2017}, dune formation \citep{kleinhans_sorting_2004}, and fault slip \citep{siman-tov_gravity-independent_2018,itoh_geological_2019}.
Recent decades have seen rapid development in both physical interpretation and theoretical modeling of particle segregation, particularly in \emph{dense} granular flows \citep{gray_particle_2018,umbanhowar_modeling_2019}. However, fundamental aspects of granular segregation at the particle level remain unclear. 

Granular materials display a rich variety of segregation behaviors. In gravity-driven flows, small particles tend to percolate downward under gravity through interstices between large particles, displacing large particles upward. By contrast, systematic evidence of reverse segregation (i.e., large particles sink) has been reported depending on the size and density ratios of particle species, as well as the species concentration \citep{thomas_reverse_2000,felix_evidence_2004}. In the absence of gravity (e.g., lateral segregation in vertical silo flows), large particles tend to migrate toward high shear rate regions \citep{fan_phase_2011,itoh_geological_2019}, but the tendency reverses when the flow becomes dilute \citep{fan_phase_2011}. Although different mechanisms including geometric effects \citep{savage_particle_1988}, mass effects \citep{felix_evidence_2004}, and shear gradient dependence \citep{fan_phase_2011} have been proposed, a unified picture remains elusive. As a result, current theoretical predictions rely on \emph{ad hoc} assumptions or phenomenological closures \citep{gray_theory_2005,marks_grainsize_2012,hill_segregation_2014,fan_modelling_2014,schlick_modeling_2015,larcher_evolution_2015,tunuguntla_comparing_2016,chassagne_discrete_2020}.

The single intruder particle limit provides an avenue to investigate the physics of granular segregation. Previous studies using this approach focused on segregation kinematics \citep{tripathi_numerical_2011,van_der_vaart_underlying_2015,jing_micromechanical_2017} and forces \citep{tripathi_numerical_2011,guillard_scaling_2016,van_der_vaart_segregation_2018,staron_rising_2018,kumar_theoretical_2019,jing_rising_2020}. \citet{guillard_scaling_2016} were the first to propose and use a virtual spring-based ``force meter'' in numerical simulations, which allows measurement of the segregation force on single intruder particles in sheared granular beds. In this approach, an intruder particle is tethered to a virtual spring that acts only in the segregation ($z$) direction (perpendicular to the shear flow in the $x$-direction). The spring applies a restoring force on the intruder that opposes segregation. This restoring force can be used to determine the segregation force $F_{seg}$, which, by definition, is nothing more than the net contact force in the $z$-direction on the intruder due to particle-particle interactions. It is therefore convenient, and unambiguous, to view segregation as a result of the imbalance between $F_{seg}$ and other forces such as the gravitational force (if present). The central goal of this work is to characterize $F_{seg}$ as a function of local flow conditions for various particle properties and system flow parameters, and validate this description across a wide range of different flow geometries.

\citet{guillard_scaling_2016} showed that, in two-dimensional (2D) confined flows, $F_{seg}$ can be expressed as two additive terms that scale with the local pressure gradient ($\partial p/\partial z$) and the local shear stress gradient ($\partial \tau/\partial z$),

\begin{equation}\label{eq:scaling1}
	F_{seg} \Big| _\textrm{confined flow} =
	-\mathcal{A}(\mu,R)\frac{\partial p}{\partial z}V_i +
	\mathcal{B}(\mu,R)\frac{\partial \tau}{\partial z}V_i,
\end{equation}

\noindent where $\mathcal{A}(\mu,R)$ and $\mathcal{B}(\mu,R)$ are dimensionless functions, $\mu=\tau/p$ is the local effective friction, $R$ is the intruder-to-bed particle size ratio, and $V_i$ is the intruder volume. Note that Guillard \emph{et al}.'s original expression was applied to 2D disks and given in terms of the intruder ``area'' instead of volume. Note also that here ``pressure'' and ``normal stress'' are interchangeable ($p=\sigma_{zz}$) because small differences in normal stress components ($\sigma_{xx}$, $\sigma_{yy}$, $\sigma_{zz}$) are neglected, and $-\partial{p}/\partial{z}=\phi\rho g_z$ indicates (positive) hydrostatic pressure gradients, where $\phi$ is the bulk packing fraction, $\rho$ is the material density of bed particles, and $g_z$ is the $z$-component of the gravitational acceleration. 
% Although (\ref{eq:scaling1}) has been briefly compared to results of free surface flow down inclines \citep{guillard_scaling_2016}, its validity beyond 2D confined flows remains largely unexplored.
% Following the $\mu(I)$ rheology, where $I=\dot{\gamma}d/\sqrt{p/\rho}$ is the inertial number ($d$ and $\rho$ are the bed particle size and density, respectively), the influence of other local flow properties (e.g., pressure $p$, shear rate $\dot{\gamma}$) are encoded in the dependence of $f_n$ and $f^s$ on $\mu$.

 Expression (\ref{eq:scaling1}) describes gravity- and shear-driven segregation in \emph{confined}, wall-driven flows and has inspired follow-up studies including \citet{van_der_vaart_segregation_2018} and \citet{jing_rising_2020} for different flow geometries. However, several important questions remain unexplored. First, although the two terms in (\ref{eq:scaling1}) appear to separate normal- and shear-stress gradient contributions, the dependence of both pre-factors $\mathcal{A}$ and $\mathcal{B}$ on $\mu$ indicates that the two effects remain coupled, since $\mu$ depends on both $p$ and $\tau$, and, hence, their gradients. The influence of shear stress profiles seems to be unclear in other geometries as well. For example, in three-dimensional (3D) inclined chute flows, \citet{van_der_vaart_segregation_2018} showed that the total segregation force is insensitive to shear stress gradients (which vary with the chute inclination), a finding we confirmed using a controlled-velocity approach that allows shear stress profiles to be specified \citep{jing_rising_2020}. Second, while it is generally accepted that the pressure gradient-induced segregation force is related to ``granular buoyancy'' \citep{guillard_scaling_2016,van_der_vaart_segregation_2018,jing_rising_2020}, the physical origin of the shear stress gradient contribution remains unexplained \citep{guillard_scaling_2016}.
 % Indeed, from a continuum modeling perspective, because the shear stress gradient $\partial \tau/\partial z$ (specifically, $\partial\tau_{zx}/\partial{z}$) does not appear in the $z$-momentum equation of the flow, it is unclear how it might contribute to the segregation force in the $z$-direction. Therefore, the $\partial \tau/\partial z$ dependence is presumably a proxy for a more fundamental mechanism.
 Third, as noted by \citet{guillard_scaling_2016}, the scaling law described by (\ref{eq:scaling1}) is based on 2D \emph{confined} flows and does not predict the sinking of very large intruders observed in \emph{free surface} flow experiments \citep{felix_evidence_2004}, which raises the question of how (\ref{eq:scaling1}) applies to 3D unconfined flow configurations.

We recently developed a scaling law for $F_{seg}$ that predicts whether an intruder rises or sinks in free surface flows \citep{jing_rising_2020}, matching extensive experimental results across the broad size-density parameter space explored by \citet{felix_evidence_2004} and agreeing with inclined chute flow simulation results, including those of \citet{van_der_vaart_segregation_2018}. The scaling law has a simple, buoyancy-like form:

\begin{equation}\label{eq:scaling2}
	F_{seg} \Big| _\textrm{free surface flow} \approx F_{seg} \Big| _\textrm{``linear'' flow} =	-f(R)\frac{\partial{p}}{\partial{z}}V_i,
\end{equation}

\noindent where $f(R)$ is dimensionless and the flow velocity is ``linear'' (elaborated below). In contrast to $\mathcal{A}(\mu,R)$ and $\mathcal{B}(\mu,R)$ expressions in (\ref{eq:scaling1}), $f(R)$ is insensitive to local flow properties (e.g., $\mu$). This finding indicates that $F_{seg}$ depends only on pressure gradients, but not shear stress gradients such as the second term of (\ref{eq:scaling1}), in free surface flows that have an approximately linear velocity profile. Indeed, the buoyancy-like scaling of (\ref{eq:scaling2}) on its own captures the chute flow results of \cite{van_der_vaart_segregation_2018},  although they further decompose $F_{seg}$ into separate lift- and buoyancy-like components.

The scaling law (\ref{eq:scaling2}) is based on controlled-velocity flows where a stabilizing algorithm enforces a linear velocity profile (i.e., constant shear rate); gravity is also included to introduce inhomogeneous pressure and shear stress profiles \citep{jing_rising_2020}. This linear-velocity flow represents an elementary flow where the segregation force can be easily connected with local flow properties, such as the shear rate, pressure, and stress gradients. A linear velocity profile is an accurate approximation for free surface flows at least over the extent of the profile in the vicinity of an intruder particle. However, for wall-driven flows where the velocity profile can be highly nonlinear, the local shear rate can vary significantly over a distance comparable to the intruder size \citep{fan_phase_2011,guillard_scaling_2016}. In these cases, higher-order effects may occur and the linear velocity assumption is not always appropriate.

To extend the applicability of the scaling (\ref{eq:scaling2}) to more general situations where the velocity profile may be nonlinear, we propose in this paper that an additional contribution to the segregation force is associated with the local curvature of the velocity profile (i.e., the shear rate gradient $\partial \dot\gamma/\partial z$, where $\dot\gamma=\partial{u}/\partial{z}$ and $u$ is the flow velocity in the streamwise $x$-direction). A unified form for $F_{seg}$ is proposed:

\begin{equation}\label{eq:scaling}
	F_{seg} \Big|_\textrm{unified} =
	-f^g(R)\frac{\partial{p}}{\partial{z}}V_i +
	f^k(R)\frac{p}{\dot{\gamma}}\frac{\partial \dot{\gamma}}{\partial z}V_i,
\end{equation}

\noindent where the first term is identical to (\ref{eq:scaling2}) and is gravity induced (hence $f^g(R)$; note that we only consider pressure gradients induced by gravity, although rotation or other body forces can also induce pressure gradients), while the second term represents a kinematics contribution (hence $f^k(R)$) that is related to the curvature of the velocity profile. Functional forms of $f^g(R)$ and $f^k(R)$ are established below as expressions (\ref{eq:fg}) and (\ref{eq:fk}), respectively. It is important to note that, as demonstrated below, both $f^g(R)$ and $f^k(R)$ are independent of $\mu$ and the kinematics contribution is universal for all flow geometries that we consider, including confined and free surface flows, for a wide range of flow conditions from quasistatic to inertial. Interestingly, the kinematics description (\ref{eq:scaling}) and the stress description (\ref{eq:scaling1}) are approximately equivalent if (and only if) the flow obeys a \emph{local rheology} \citep{forterre_flows_2008}, i.e., shear stresses depend only on local shear rates, leading to the $\mu$ dependence in (\ref{eq:scaling1}); see Appendix~\ref{appA}. However, this equivalence breaks down in flow regions that exhibit a \emph{nonlocal rheology} \citep{kamrin_non-locality_2019}, thus limiting the generality of (\ref{eq:scaling1}). Lastly, the specific form of the second term in (\ref{eq:scaling}) is inferred from a dimensional argument, and the relevance of all parameters ($p$, $\dot\gamma$, $\partial \dot\gamma/\partial z$, and $V_i$) is verified in this paper based on a comprehensive parametric study. In particular, while $f^g(R)$ is studied and established in our recent work \citep{jing_rising_2020}, $f^k(R)$ is developed in this paper by extending the controlled-velocity flow from constant shear rate \citep{jing_rising_2020} to constant shear rate gradient (i.e., controlled curvature of the velocity profile), and varying the curvature extensively.

In \S\ref{sec:methods} we introduce the simulation scheme that allows flow kinematics to be arbitrarily controlled, as well as other flow geometries that we use for validation. Then, the approach to measuring $F_{seg}$ and model details are presented. In \S\ref{sec:results} we first focus on $R=2$ and characterize the kinematics contribution to $F_{seg}$ in the absence of gravity, after which we introduce gravity and show that the two terms in (\ref{eq:scaling}) are indeed additive. The proposed scaling law is then compared with results from other geometries, highlighting the universality of gravity- and kinematics-induced segregation forces. Finally, results for varying $R$ are presented to establish the $R$ dependence in scaling law (\ref{eq:scaling}). Conclusions are drawn in \S\ref{sec:conclusion}.

\section{Methods}\label{sec:methods}

\subsection{Flow configurations}
\begin{figure}
  \centerline{\includegraphics[width=0.9\linewidth]{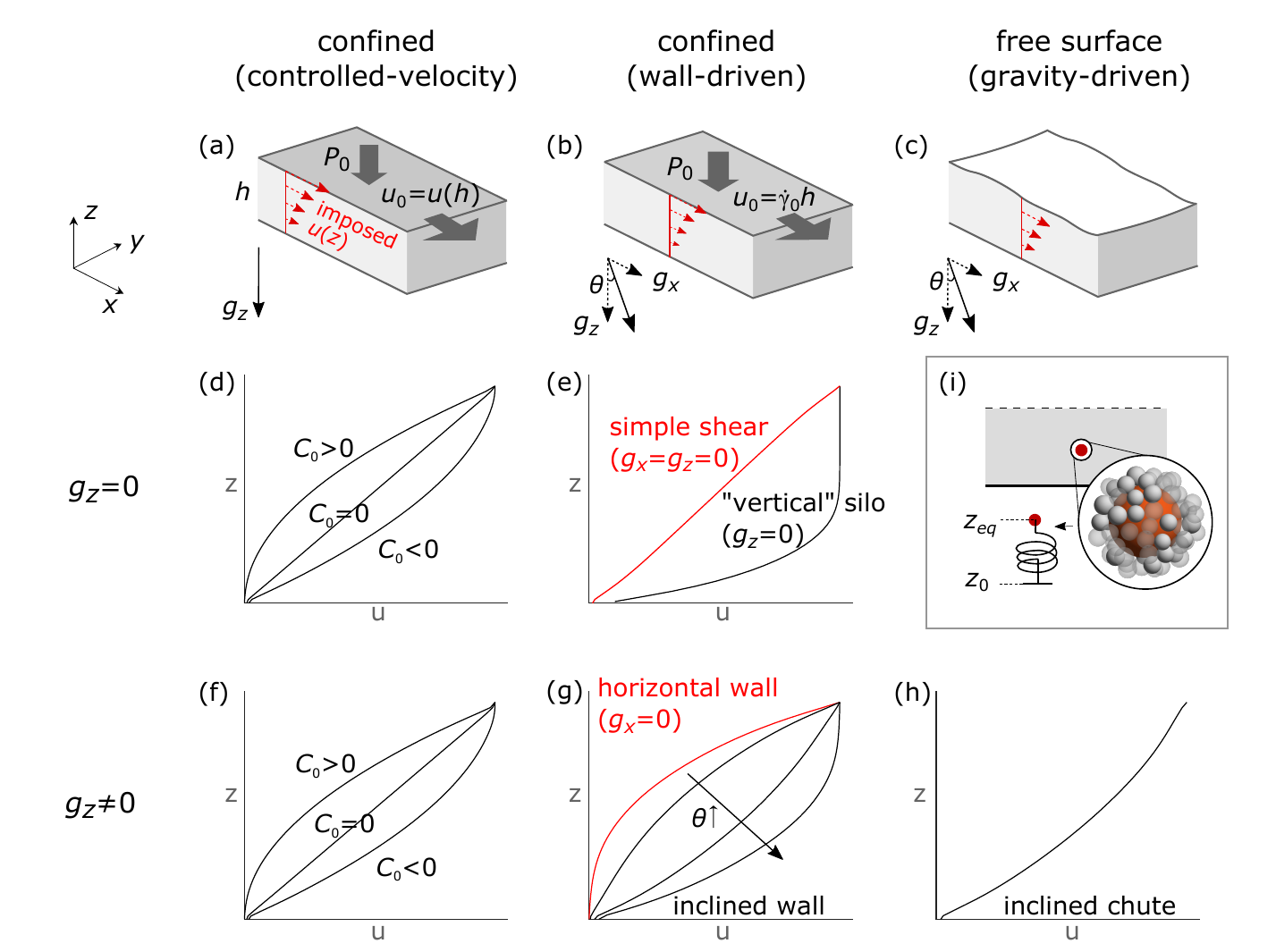}}% Images in 100% size
  \caption{(a--h) Flow configurations and associated velocity profiles (see text for details). (i) Intruder particle (red) in the flow and tethered to a virtual vertical spring for segregation force measurement.}
\label{fig:setup}
\end{figure}

We use the open-source discrete element method (DEM) code \textsc{liggghts} \citep{kloss_models_2012} to simulate several different dense granular flows, which can be classified broadly into ``confined'' and ``free surface'' flows (figure~\ref{fig:setup}). For confined flows, we vary the velocity profile with two different approaches: first, by directly controlling the velocity field (referred to as ``controlled-velocity'' flows; see below); second, by varying the direction of gravity in ``wall-driven'' flows. In each flow, a single intruder particle is placed in the middle of the flow depth to measure the segregation force $F_{seg}$ (see figure~\ref{fig:setup}i and \S\ref{sec:methods_force}). The focus is on how $F_{seg}$ depends on the local curvature of the velocity profile.
% which may vary significantly depending on the specific boundary conditions and external forcing applied in each flow.

As shown in figure~\ref{fig:setup}, periodic boundaries are imposed in the streamwise ($x$) and spanwise ($y$) directions and the domain is sufficiently wide in both directions that, in steady state, gradients of the flow occur only in the normal ($z$) direction; that is, we only consider segregation in the $z$-direction. For controlled-velocity flows, gravity is always normal to the bottom wall, while for other flows, gravity may be tilted in the $xz$-plane to an angle of $\theta$ with respect to the $z$-axis. The gravity vector is denoted as $(g_x,0,g_z)=(g\sin\theta,0,-g\cos\theta)$, where $g$ is the magnitude of the gravitational acceleration.
% A flow with $g_z=0$ does not have gravity in the direction of segregation (simply referred to as ``no gravity'') and segregation is only driven by shear. When $g_z\ne0$, pressure gradients develop in the $z$-direction and contribute to segregation.
To aid interpretation of the segregation direction, we use ``up'' and ``down'' (or similar terms) to refer to the positive and negative directions of the $z$-axis consistent with that in figure~\ref{fig:setup}, even when gravity is turned off or is parallel to the flow (i.e., $g_z=0$). The same convention applies to ``top'' and ``bottom'' walls. Bottom walls are always immobile, while top walls (absent in free surface flows) are reactive in the $z$-direction to maintain an overburden pressure $P_0$ and translate with velocity $u_0$. All walls are roughened by randomly distributed stationary particles to reduce slippage \citep{jing_characterization_2016}.
% In controlled-velocity flows, the top wall moves in the $x$-direction following the imposed velocity profile (see below). In confined flows, velocity is imposed at the top wall to drive the flow. Free surface flows are driven simply by tilting gravity.

In the following, each flow type is described with an emphasis on the curvature of its velocity profile and how the curvature is systematically varied.

\subsubsection{Confined, controlled-velocity flows}

Controlled-velocity flow, in which the flow velocity profile is specified, has been used previously to study granular rheology. While linear velocity profiles are usually imposed in the absence of gravity to achieve homogeneous shear \citep{lerner_unified_2012,clark_critical_2018}, it is also possible to impose arbitrary velocity profiles without gravity \citep{saitoh_nonlocal_2019} or to add gravity for linear velocity profile flows \citep{fry_effect_2018,duan_segregation_2020}. Controlling the velocity profile allows us to vary shear rate gradients (i.e., curvatures of the velocity profile) independently of gravity, thereby isolating the gravity- and kinematics-related contributions to the segregation force.

To impose a particular streamwise velocity profile $u(z)$, a stabilizing force (in the $x$-direction) is applied to each particle at each DEM time step \citep{fry_effect_2018}, including the intruder particle, and the top wall is translated at a matching speed $u(h)$, where $h$ is the flow thickness. The stabilizing force applied on a particle is $A(u(z_p)-u_p)$, where $z_p$ and $u_p$ are the instantaneous particle position and velocity, respectively, and $A$ is a constant. We use $A=0.1$~\si{N\cdot s/m} in the controlled-velocity flows below, but increasing or decreasing $A$ by an order of magnitude does not change the results significantly. Indeed, as detailed below, controlled-velocity flows follow the same granular rheology as homogeneous shear flow, indicating that particle-particle collisions dominate the particle rheology even when stabilizing forces are imposed.
% At the continuum level, applying stabilizing forces is equivalent to adding a force term to the $x$-momentum equation of the flow, but the profile of this force term is chosen by particle dynamics ().

To generate a constant shear rate gradient flow, we specify $u(z)$ as

\begin{equation}\label{eq:uz}
	u(z) = (\dot\gamma_0-\frac{C_0h}{2})z+\frac{C_0}{2}z^2,
\end{equation}

\noindent where $\dot\gamma_0$ and $C_0$ are, respectively, a characteristic shear rate and the controlled curvature (i.e., shear rate gradient). For this profile, we have

% \begin{subeqnarray}
%   \dot\gamma(z) & = & \dot\gamma_0-\frac{C_0h}{2}+C_0z,\\[3pt]
%   \frac{\partial\dot\gamma}{\partial z}(z) & = & C_0.
% \end{subeqnarray}

\begin{equation}\label{eq:gamz}
	\dot\gamma(z) = \dot\gamma_0-\frac{C_0h}{2}+C_0z,
\end{equation}

\noindent and

\begin{equation}\label{eq:dgamz}
	\frac{\partial\dot\gamma}{\partial z} = C_0.
\end{equation}

\noindent The specific form of (\ref{eq:uz}) is chosen for several reasons. First, the local shear rate in the middle of the flow (where the intruder is placed) is $\dot\gamma(h/2)=\dot\gamma_0$ independent of $C_0$. Second, the shear rate gradient in the simulation is homogeneous (except within a few bed particle diameters of the upper and lower boundaries; see figure~\ref{fig:typical} below) based on (\ref{eq:dgamz}); by varying $C_0$ from negative to positive, the concavity and local curvatures of the velocity profile around the intruder are systematically varied and precisely controlled (figures~\ref{fig:setup}d and f). Third, the velocity at the upper boundary, $z=h$, is $u_0=\dot\gamma_0h$, which is consistent with the top wall velocity in wall-driven flows (figure~\ref{fig:setup}b). Furthermore, we constrain gravity so that it is always in the $z$-direction; that is, $g_x=0$.  Other forms of $u(z)$, such as power-law and exponential functions, could be used in place of (\ref{eq:uz}), but these do not have the advantage of a constant $\partial\dot\gamma/\partial{z}$ throughout the flow domain. Nevertheless, we have verified that these alternative velocity profiles produce segregation forces that are consistent with the scaling (\ref{eq:scaling}).

% use power-law velocity profiles in order to conveniently change the concavity by varying the power index $n$. The velocity profile follows
% \begin{equation}\label{eq:powerlaw}
% 	u(z) = \frac{1}{nz_0^{n-1}}\dot{\gamma}_0z^n 
% 	% \propto \dot{\gamma}_0z^n
% \end{equation}
% \noindent where $\dot{\gamma}_0$ is a characteristic shear rate imposed at a given location $z_0$ (typically the middle of the flowing layer, around which measurement takes place). This is to study the effect of local curvatures (i.e., the shear rate gradient) while controlling the local shear rate. As shown in figure~\ref{fig:setup}d, $n=1$ represents a linear velocity profile (with a constant shear rate $\dot{\gamma}_0$), while $n<1$ and $n>1$ lead to concave-up (negative-curvature) and concave-down (positive-curvature) velocity profiles, respectively.

% When $n\ne1$, the specific form we use is
% \begin{equation}
% 	u(z) = \frac{1}{nz_0^{n-1}}\dot{\gamma}_0z^n
% \end{equation}
% \noindent so that a shear rate $\dot{\gamma}_0$ is imposed at a given location $z_0$ (typically the middle of the flowing layer, around which main measurement takes place). 

Apart from systematically varying the shear rate gradient $C_0$, we also vary $P_0$, $\dot{\gamma}_0$, and $g_z$ to explore their effects on the segregation force on a single intruder particle. A key advantage of the controlled-velocity geometry is that the imposed velocity profile is unaffected by gravity; the flow adjusts its shear stress profiles in response to changed gravitational fields while maintaining the same flow kinematics. As shown schematically in figures~\ref{fig:setup}d and f, identical velocity profiles are achieved for the $g_z=0$ and $g_z\ne0$ cases while keeping other system parameters constant.

\subsubsection{Wall-driven flows}

In wall-driven flows, overburden pressure $P_0$ and velocity $u_0=\dot\gamma_0h$ are imposed at the top wall to drive the flow without directly controlling the velocity profile (note that in wall-driven flows we use $\dot\gamma_0$ to characterize the top wall velocity, consistent with the notation for controlled-velocity flows). The concavity of the velocity profile of wall-driven flows is altered by varying the direction of gravity (figures~\ref{fig:setup}e and g). With no gravity ($g_x=g_z=0$), the flow is simple shear with a nearly linear velocity profile (figure~\ref{fig:setup}e), and segregation does not occur because both the pressure gradient and shear rate gradient are zero. 

When gravity is parallel to the flow direction ($\theta=90^\circ$, $g_x\ne0$, $g_z=0$), shear stress gradients develop along the $z$-direction and the velocity profile is concave up (negative $\partial\dot\gamma/\partial{z}$), as shown in figure~\ref{fig:setup}(e). Shear is localized near the bottom wall, above which is a plug-flow zone. This flow is similar to half of a ``vertical'' silo flow \citep{gdr_midi_dense_2004}, albeit horizontally placed in our coordinate system. Segregation in the $z$-direction is driven only by shear as there is no pressure gradient ($g_z=0$).

When gravity is perpendicular to the flow ($\theta=0$, $g_x=0$, $g_z\ne0$), a flowing layer develops near the top wall with concave-down (positive $\partial\dot\gamma/\partial{z}$) velocity profiles (figure~\ref{fig:setup}g). Shear stress is homogeneous in this geometry, because the only external forcing in the $x$-direction is applied from the top wall \citep{guillard_scaling_2016}. However, both the pressure gradient due to gravity and the non-linear velocity profile (or, the shear rate gradient) are expected to contribute to segregation.

As $\theta$ increases from $0$ toward $90^\circ$ ($\theta>0$, $g_x\ne0$, $g_z\ne0$; ``inclined wall-driven'' flows in figure~\ref{fig:setup}g), the velocity profile changes from concave down to concave up, and the kinematics- and gravity-related segregation mechanisms can either compete or cooperate. 

\subsubsection{Free surface, gravity-driven flows}

Common free surface flows include chute flow, heap flow, and surface flow in rotating tumblers. Here we study relatively thick flows (about $40$ particles deep) down an inclined streamwise and spanwise periodic chute that exhibit Bagnold-like, concave-up velocity profiles (figure~\ref{fig:setup}h). Thin chute flows \citep{silbert_granular_2003,louge_model_2003,weinhart_closure_2012,kamrin_nonlocal_2015} or shallow flowing layers in heap and rotating-drum flows \citep{gdr_midi_dense_2004,kamrin_nonlocal_2012} will be addressed in separate work as these flows are likely to be strongly affected by bottom or sidewall boundaries.
% Nevertheless, similar rise-sink transitions have been observed in all three free-surface configurations \citep{felix_evidence_2004,jing_rising_2020}; extension from thick chute flows to thin flows will be addressed in the future.

\subsection{Segregation force measurement}\label{sec:methods_force}

We measure the segregation force on a single intruder particle in each flow simulation following the approach of \citet{guillard_scaling_2016}. The intruder is tethered to a virtual spring that senses forces only in the $z$-direction, which allows the intruder to deviate from the initial height $z_0$ and fluctuate around an equilibrium position $z_{eq}$ (figure~\ref{fig:setup}i). In equilibrium, $F_{seg}$, the net contact force on the intruder perpendicular to the flow (in the $z$-direction), is balanced by the spring force and the intruder weight, i.e., $F_{seg}=k(z_{eq}-z_0)+m_ig_z$, where $k$ is the spring stiffness and $m_i$ is the intruder mass. Note that $F_{seg}$ represents the \emph{mean} segregation force, even though the random action of contacting particles fluctuates in time. The uncertainty in $F_{seg}$ (error bars) is estimated based on temporally correlated fluctuations of the intruder position around $z_{eq}$ \citep{zhang_calculation_2006}. The measurement of $F_{seg}$ is insensitive to $k$ for 3D configurations \citep{van_der_vaart_segregation_2018} as the intruder is free to explore the $xy$-plane. We use stiff springs (typically, $k=100$~\si{N/m}) to ensure that $z_{eq}$ is close to $z_0$ such that local flow conditions around the intruder can be a priori controlled (or estimated in flows without directly controlled velocity profiles).

\subsection{Model parameters and flow conditions}\label{sec:methods_para}

The flow domain in all simulations is $30d$ long ($x$), $20d$ to $30d$ wide ($y$), and $40d$ deep ($z$) (adjusted to avoid boundary effects), and contains bed particles of diameter $d$ and density $\rho$. A single intruder particle of size ratio $R=d_i/d$ and density ratio $R_\rho=\rho_i/\rho$, where $d_i$ and $\rho_i$ are the intruder diameter and density, is placed in the middle of the flow. In previous work we varied both $R$ and $R_\rho$ to study forces driving combined size and density segregation \citep{jing_rising_2020}. Here, to simplify the parameter space, we only report results for $d=5$~\si{mm} (with $10\%$ size polydispersity), $\rho=2500$~\si{kg/m^3}, $R_\rho=1$, and $0.2\leqslant R \leqslant8$. However, varying $d$, $\rho$, or $R_\rho$ does not change the scaling of $F_{seg}$. Particle interactions are calculated using the Hertz contact model with Young's modulus $5\times10^7$~\si{Pa}, Poisson's ratio $0.4$, restitution coefficient $0.8$, and friction coefficient $0.5$; varying these parameters has negligible influence on the results, except for friction $\lesssim 0.3$ (see Supplemental Material for \citet{jing_rising_2020}).

System parameters are varied for each flow geometry (figure~\ref{fig:setup}) to achieve a wide range of local flow conditions (e.g., $\partial{p}/\partial{z}$, $\partial{\dot\gamma}/\partial{z}$, $p$, $\dot\gamma$) around the intruder, which are then associated with $F_{seg}$ according to the proposed relation (\ref{eq:scaling}). Local flow conditions are estimated for each simulation at $z=z_{eq}$ based on spatially and temporally averaged flow fields along the $z$-direction. Steady-state flow and stress profiles are estimated based on $1d$-thick bins along the flow depth ($z$) that span the simulation domain in the $xy$-plane. For a given instant at steady state, we first compute averaged velocity and contact stresses \citep{silbert_granular_2001} in each bin based on particles centered in that bin, including the intruder particle, and then smooth the depth-wise profile spatially using a moving average filter (typically spanning five equally weighted bins). The profiles are then averaged in time, typically using $200$ snapshots for steady state conditions. First or second-order gradients of the velocity and stress profiles (e.g., $\dot\gamma$, $\partial{\dot\gamma}/\partial{z}$, $\partial{p}/\partial{z}$) are calculated using central differences. Although an intruder particle can change local flow structures in its vicinity \citep{van_der_vaart_segregation_2018,jing_rising_2020}, we have verified that local disturbances due to the presence of the intruder are smoothed out (see smoothed flow profiles in figures~\ref{fig:typical}, \ref{fig:typical_g}, and \ref{fig:wall_chute} below) and that details of the averaging method do not affect the results.

The flow at $z=z_{eq}$ is characterized by the local inertial number $I(z_{eq})=\dot\gamma(z_{eq})d\sqrt{\rho/p(z_{eq})}$, which is varied broadly from $0.004$ to $0.44$ with $500\leqslant P_0 \leqslant 2500$~\si{Pa} and $10\leqslant\dot\gamma_0\leqslant40$~\si{s^{-1}} for controlled-velocity and confined flows or $22^\circ\leqslant\theta\leqslant28^\circ$ for inclined chute flows; varying $g$ also influences $I(z_{eq})$. For brevity, we mainly report $I(z_{eq})$ (simply referred to as $I$ below) instead of system parameters when identifying flow characteristics; detailed simulation parameters (both controlled system parameters and measured local parameters) are provided as supplementary material.

\section{Results and discussion}\label{sec:results}

\subsection{Segregation forces for $R=2$}

Focusing first on size ratio $R=2$ and density ratio $R_\rho=1$, we vary local curvatures in controlled-velocity flows following (\ref{eq:uz}) in the absence of gravity ($g_z=0$) to study the kinematics-dependent part of $F_{seg}$, and then add gravity ($g_z>0$) to study the gravity-dependent part. Results from other flow geometries are used to validate the  scaling.

\subsubsection{Kinematics contribution (no gravity)}

\begin{figure}
  \centerline{\includegraphics[width=0.9\linewidth]{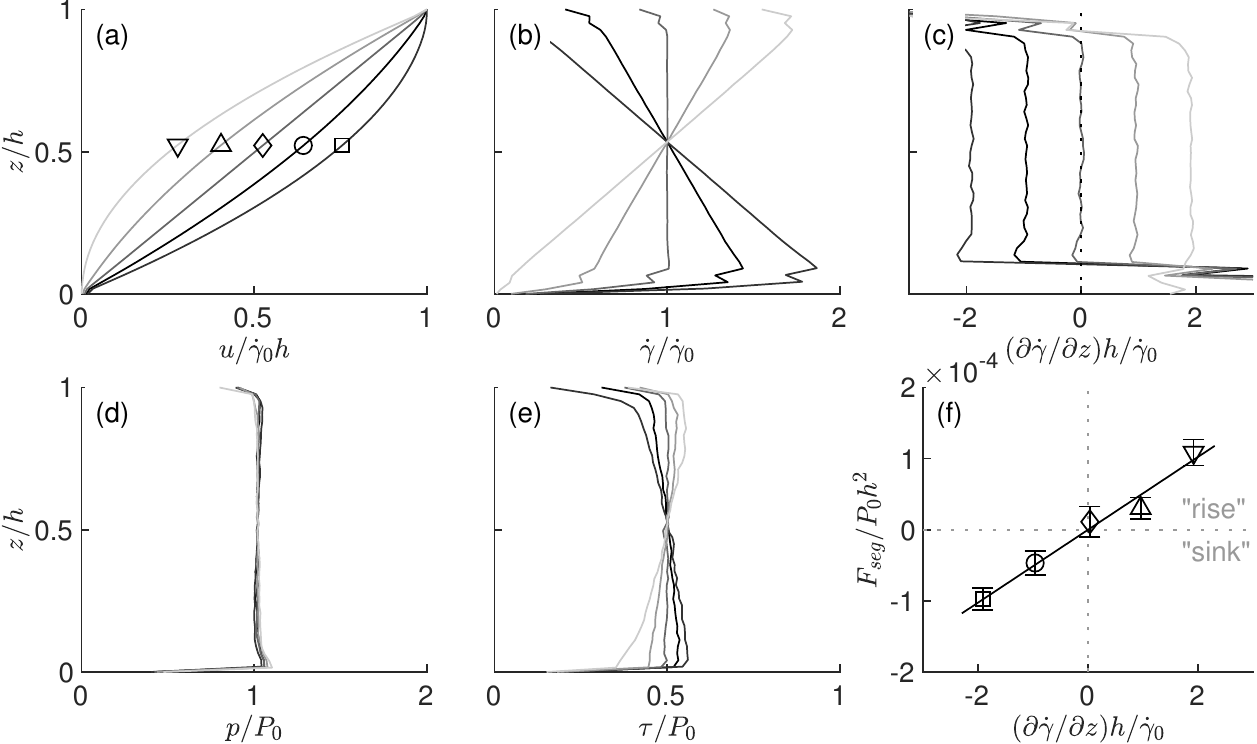}}
  \caption{
  Typical results for controlled-shear-rate-gradient flows ($g_z=0$, $P_0=1000$~\si{Pa}, $\dot{\gamma_0}=30$~\si{s^{-1}}). (a--e) Profiles of the normalized velocity, shear rate, shear rate gradient, pressure, and shear stress, respectively. Darker to lighter colors indicate $C_0$ varying from negative to positive ($-2\leqslant C_0h/\dot\gamma_0\leqslant2$). Symbols in (a) indicate the steady-state $z$-location of the intruder. (f) Measured dimensionless segregation force $F_{seg}/P_0h^2$ vs.\ dimensionless shear rate gradient $(\partial\dot\gamma/\partial{z})h/\dot\gamma_0$. Symbols correspond to those in (a), and the solid line indicates a linear fit to the data through the origin. Error bars represent the uncertainty of $F_{seg}$ due to its fluctuations in time (see \S\ref{sec:methods_force}).
       }
\label{fig:typical}
\end{figure}

Figure~\ref{fig:typical} shows data from a representative set of controlled-velocity simulations with varying $C_0$ but fixed $P_0$ and $\dot{\gamma}_0$ (thus fixed $I$ around the intruder); $\dot\gamma_0$, $P_0$, and $h$ are used for normalization. In figure~\ref{fig:typical}(a), as $C_0$ is varied from negative (darker curves) to positive (lighter curves), the velocity profile varies from concave up to concave down, including a linear case with no curvature ($C_0=0$). A symbol marks the intruder position $z_{eq}$ on each curve, which is alway midway between the upper and lower walls to minimize wall effects. Figure~\ref{fig:typical}(b) shows that, although the shear rate profile varies with $C_0$, the local shear rate at the position of the intruder ($z/h\approx0.5$) is always $\dot{\gamma}_0$ due to the imposed velocity profile (\ref{eq:uz}). Figure~\ref{fig:typical}(c) shows that the imposed velocity profile results in a constant local shear rate gradient $\partial\dot\gamma/\partial{z}$ across nearly the entire flow domain (except at the upper and lower walls) with dimensionless curvatures $C_0h/\dot\gamma_0$ varying from $-2$ to $2$. Since $g_z=0$, pressure is uniform across the flow domain ($p(z)/P_0\approx1$), as shown in figure~\ref{fig:typical}(d). However, the shear stress in figure~\ref{fig:typical}(e) varies somewhat with depth due to the imposed velocity profiles, although its value at the location of the intruder is identical for all cases.

The key measurement here, the dimensionless force $F_{seg}/P_0h$ exerted by bed particles on the intruder based on the measured virtual spring force, is plotted in figure~\ref{fig:typical}(f) against the dimensionless local curvature $(\partial\dot\gamma/\partial{z})h/\dot\gamma_0$. It is evident that a non-linear flow velocity profile alone induces a net contact force on the intruder that drives segregation. Since $F_{seg}$ is the only (net) force acting on the intruder in the no-gravity situation, negative values of $F_{seg}$ correspond to the large intruder ``sinking'' (in the coordinate system of figure~\ref{fig:setup}) toward high-shear regions, consistent with the trend of shear-driven segregation in \emph{dense} vertical silo flows \citep{fan_phase_2011}. Similarly, positive values of $F_{seg}$ indicate the large intruder ``rising,'' again, toward high-shear regions (see ``rise'' and ``sink'' in figure~\ref{fig:typical}f). The relationship between $F_{seg}$ and $\partial\dot\gamma/\partial{z}$ is linear and through the origin, indicating no segregation force when there is no shear rate gradient, as would be expected. Note that both negative and positive curvatures are considered here despite the apparent symmetry of the segregation behavior because the flow system we use is slightly asymmetric; the bottom wall is fixed whereas the top wall moves slightly in the $z$-direction in response to the constant $P_0$ boundary condition.

\begin{figure}
  \centerline{\includegraphics[width=0.9\linewidth]{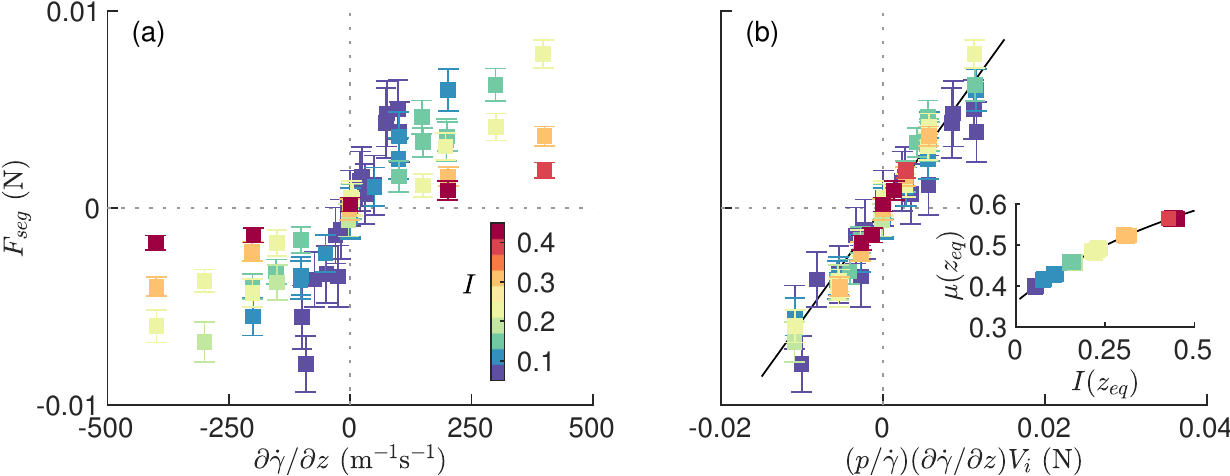}}% Images in 100% size
  \caption{ 
  Influence of $P_0$ and $\dot\gamma_0$ on $F_{seg}$ in controlled-velocity flows ($g_z=0$). A range of local inertial numbers, $0.05\leqslant I \leqslant 0.44$ (color bar), is explored with $500\textrm{ Pa}\leqslant P_0\leqslant2000\textrm{ Pa}$ and $10\textrm{ s}^{-1}\leqslant \dot\gamma_0 \leqslant 40\textrm{ s}^{-1}$; for each ($P_0$, $\dot\gamma_0$) combination, $C_0h/\dot\gamma_0$ is varied from $-2$ to $2$ to generate negative to positive curvatures $\partial\dot\gamma/\partial{z}$. (a) Non-collapse of $F_{seg}$ vs.\ $\partial\dot\gamma/\partial{z}$. (b) Collapse of $F_{seg}$ vs.\ $(p/\dot{\gamma})(\partial\dot\gamma/\partial{z})V_i$. The solid line is a linear fit through ($0,0$) with slope $0.57$. Inset: Local $\mu(I)$ measurements at $z=z_{eq}$. Reference curve $\mu(I)=0.36+(0.94-0.36)/(0.8/I+1)$ is obtained using our simple shear data far from boundaries and with no intruder.
  }
\label{fig:g0}
\end{figure}

To explore the effect of local flow conditions on $F_{seg}$, we repeat the cases in figure~\ref{fig:typical} with nine different combinations of $P_0$ and $\dot{\gamma}_0$ (in total $59$ simulations), leading to local inertial numbers $I$ varying from $0.05$ to $0.44$ (see figure~\ref{fig:g0}b inset for $\mu(I)$ data). As shown in figure~\ref{fig:g0}(a), the dependence of $F_{seg}$ on $\partial\dot\gamma/\partial{z}$ varies with $I$ when presented in physical units; the slope of the linear correlation tends to decrease as $I$ increases (from blue to red symbols). Note that since data have more scatter for small $I$ approaching the quasistatic limit ($<0.1$), a finer variation of $\partial\dot\gamma/\partial{z}$ is used in these cases resulting in more data points for small $I$.

To collapse the data, we consider the rescaled curvature from (\ref{eq:scaling}), $(p/\dot{\gamma})(\partial\dot\gamma/\partial{z})V_i$, in units of force, which results in excellent collapse for the full range of $I$ that we examine, as shown in figure~\ref{fig:g0}(b); that is,

\begin{equation}\label{eq:fcurv}
	 F^k_{seg} := F_{seg}\Big|_{g_z=0} \propto \frac{p}{\dot{\gamma}}\frac{\partial\dot{\gamma}}{\partial{z}}V_i,
\end{equation}

\noindent where $F^k_{seg}$ denotes the kinematics-induced part of $F_{seg}$. The proportionality constant is $0.57$ based on the linear fit in figure~\ref{fig:g0}(b), and the correlation passes through the origin, indicating that no other effects are present.

The force scaling in (\ref{eq:fcurv}) is based on dimensional analysis. Indeed, when $g_z=0$, natural choices for normalizing $F_{seg}^k$ and $\partial\dot\gamma/\partial{z}$ are $pd_i^2$ and $\dot\gamma/d_i$, respectively, and the resulting scaling, $F_{seg}^k/pd_i^2\propto(\partial\dot\gamma/\partial{z})d_i/\dot\gamma$, or $F_{seg}^k \propto (p/\dot\gamma)(\partial\dot\gamma/\partial{z})d_i^3$, is equivalent to (\ref{eq:fcurv}). However, here we prefer the volume-based expression (\ref{eq:fcurv}) for consistency with the buoyancy-like term in (\ref{eq:scaling}) and following the previous scaling law (\ref{eq:scaling1}). Moreover, the scaling $F_{seg}^k/pd_i^2\propto(\partial\dot\gamma/\partial{z})d_i/\dot\gamma$ is similar to the normalization in figure~\ref{fig:typical}f except that the system length scale $h$ is replaced by the intruder diameter $d_i$ (giving rise to $V_i$) and that local flow conditions are used. To confirm that $d_i$ is the relevant length scale, we verified (omitted for brevity) that varying the flow thickness $h$ does not change the scaling of $F^k_{seg}$, but doubling both $d$ and $d_i$ (with fixed $R=2$) leads to a segregation force eight times larger, as the scaling predicts. 

The curvature-based scaling (\ref{eq:fcurv}) indicates that the shear rate gradient drives segregation in the absence of gravity. Although it is also possible to express a force scale in other ways, such as one related to the shear stress gradient or the granular temperature gradient (note that no pressure gradient is present so far), we have verified that the current form (\ref{eq:fcurv}) results in the simplest scaling while other choices do not collapse the data as well as the scaling used here. For instance, using $\partial\tau/\partial{z}$ leads to scaling factors that depend on $I$ (or $\mu$), similar to those reported in \citet{guillard_scaling_2016}, which not only complicates the function but also reduces the generality of the scaling because $\mu(I)$ is not necessarily unique across flow geometries or in regions where nonlocal effects occur \citep{gdr_midi_dense_2004}; see also \S\ref{sec:valid}. Nevertheless, in Appendix~\ref{appA} we demonstrate that if the flow obeys a local rheology (e.g., $\mu(I)$), our $\partial\dot\gamma/\partial{z}$-based scaling is equivalent to the $\partial\tau/\partial{z}$-based scaling proposed by \citet{guillard_scaling_2016}.

\subsubsection{Adding gravity}

\begin{figure}
  \centerline{\includegraphics[width=0.9\linewidth]{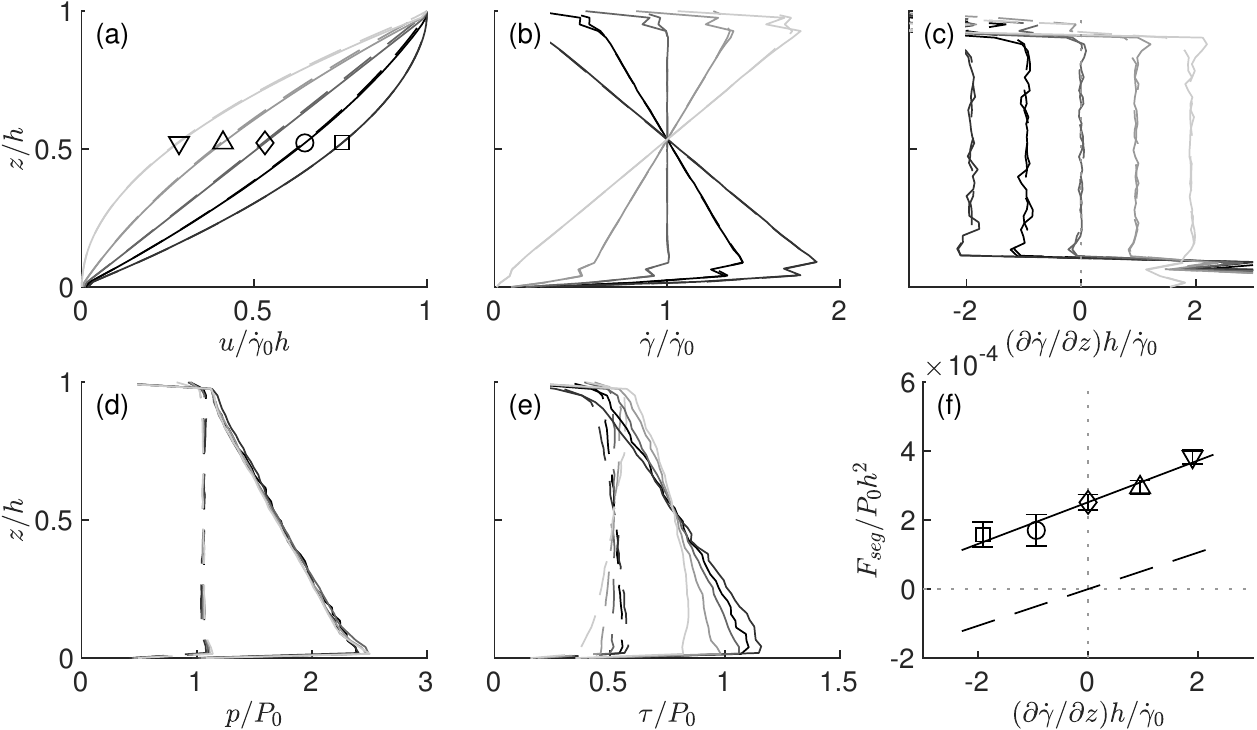}}
  \caption{ 
  Same as figure~\ref{fig:typical} except with additional data for $g_z=5$~\si{m/s^2} (solid curves). Dashed curves are the $g_z=0$ data in figure~\ref{fig:typical}.
  }
\label{fig:typical_g}
\end{figure}

With gravity ($g_z>0$), $F_{seg}$ changes due to the induced pressure gradient. Figure~\ref{fig:typical_g} shows the same set of controlled-velocity flows as in figure~\ref{fig:typical} except with $g_z=5$~\si{m/s^2}. The kinematics profiles (solid curves in figures~\ref{fig:typical_g}a--c) are nearly identical to their no-gravity counterparts (dashed curves) because velocity profiles are imposed. Stress profiles, on the other hand, change significantly in response to the added gravitational field (while granular rheology remains the same; see below). Both pressure and shear stress fields in figures~\ref{fig:typical_g}(d) and (e) now include a hydrostatic component that is proportional to $\phi\rho g_z$. With these changes in the stress fields, $F_{seg}$ remains proportional to the rescaled curvature (symbols and solid line in figure~\ref{fig:typical_g}f), but with a substantial positive offset compared to the no-gravity results (dashed line in figure~\ref{fig:typical_g}f); the slopes of the two lines are nearly identical, indicating that the gravity-induced contribution to $F_{seg}$ does not change the kinematics contribution. Thus, the two terms in (\ref{eq:scaling}) are additive. 

\begin{figure}
  \centerline{\includegraphics[width=0.9\linewidth]{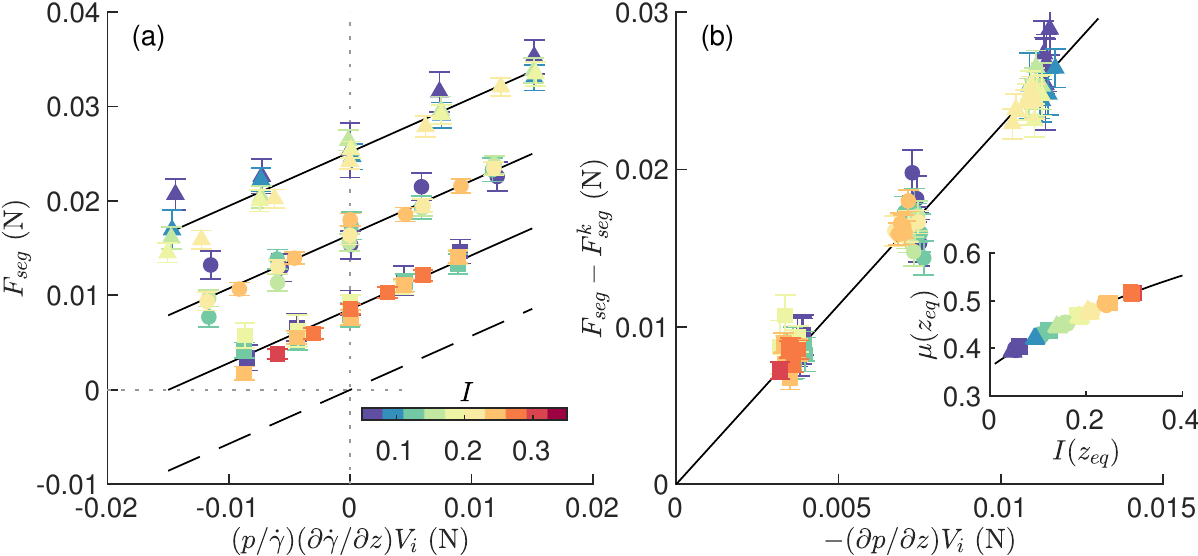}}
  \caption{
  (a) Measured segregation force in controlled-velocity flows with $g_z=\{5,9.81,15\}$~\si{m/s^2}. For each $g_z$, a range of local inertial numbers (color bar) and local curvatures (horizontal axis) are explored with $500\textrm{ Pa}\leqslant P_0\leqslant1500\textrm{ Pa}$, $10\textrm{ s}^{-1}\leqslant \dot\gamma_0 \leqslant 40\textrm{ s}^{-1}$ and $-2\leqslant C_0h/\dot\gamma_0\leqslant 2$. Solid lines are linear fits for the same $g_z$ ($g_z$ increases from bottom to top), while the dashed line represents $g_z=0$ results; all lines have the same slope of $0.57$. (b) $F_{seg}-F^k_{seg}$ vs.\ $-(\partial{p}/\partial{z})V_i$. The solid line is a linear fit with slope $2.28$ that extends through the origin. Inset: Local $\mu(I)$ measurements at $z=z_{eq}$, compared to simple shear results (curve; see figure~\ref{fig:g0} caption).
  }
\label{fig:g}
\end{figure}

Next we consider a broader range of cases than in figure~\ref{fig:typical_g} by varying $P_0$ and $\dot{\gamma}_0$ as well as $g_z$ ($75$ simulations in total). Figure~\ref{fig:g}(a) shows that the data for $F_{seg}$ at different values of $P_0$ and $\dot{\gamma}_0$, when plotted against $(p/\dot{\gamma})(\partial\dot\gamma/\partial{z})V_i$, collapse onto lines corresponding to each non-zero value of $g_z$ (similar to Figure~\ref{fig:g0}b for $g_z=0$). As $g_z$ is increased from $0$ (dashed line), the dependence of $F_{seg}$ on $p/\dot{\gamma}(\partial\dot\gamma/\partial{z})V_i$ remains linear (solid lines) with a slope independent of $g_z$ but shifted upward due to the imposed gravity. This further confirms the conclusion from figure~\ref{fig:typical_g}(f) that the two terms in (\ref{eq:scaling}) are additive. Furthermore, $F_{seg}$ remains insensitive to $I$ (indicated by the symbol colors) over the range examined ($0.05<I<0.35$; note that varying $g_z$ tends to affect the range of $I$), and the rheological data of controlled-velocity flows (when gravity is turned on) still follow the $\mu(I)$ curve for simple shear (figure~\ref{fig:g}b inset).

Because the gravity- and kinematics-related terms are additive, it is possible to use (\ref{eq:fcurv}) to characterize the gravity-induced portion of the segregation force by simply subtracting $F^k_{seg}$ from $F_{seg}$, which is plotted against $-(\partial{p}/\partial{z})V_i$ in figure~\ref{fig:g}(b). All data collapse onto a line passing through the origin, which indicates a buoyancy-like scaling consistent with the one we proposed \citep{jing_rising_2020} based only on constant-shear-rate flows (i.e., $F^k_{seg}=0$). That the fit passes through the origin indicates that $F_{seg}$ is completely described by the additive combination of gravitational and kinematic contributions, as indicated by (\ref{eq:scaling}). Note that the quadratic controlled-velocity profiles (\ref{eq:uz}) reduce to linear-velocity profiles for $C_0=0$, and using only $C_0=0$ data produces the same linear fit as that in figure~\ref{fig:g}(b) (not shown as they are virtually identical). This again supports the assumption in (\ref{eq:scaling}) that the gravity- and kinematics-induced segregation forces are additive, and indicates that the gravity-related part can be measured using linear controlled-velocity flows, as in \citet{jing_rising_2020}. Specifically,

\begin{equation}\label{eq:fgrav}
	F^g_{seg} := F_{seg} \Big|_{\partial\dot{\gamma}/\partial{z}=0} \propto -\frac{\partial{p}}{\partial{z}}V_i,
\end{equation}

\noindent where the proportionality constant is $2.28$ according to figure~\ref{fig:g}(b).
% $2.276\pm0.031$ vs $2.283\pm0.061$

\subsubsection{Validation in other flow geometries}\label{sec:valid}

In the previous two sections we establish scaling laws for the gravity- and kinematics-induced segregation forces that determine the net segregation force (for $R=2$),

\begin{equation}\label{eq:scaling_R2}
	F_{seg} = F_{seg}^g + F_{seg}^k = -f^g\frac{\partial{p}}{\partial{z}}V_i + f^k\frac{p}{\dot{\gamma}}\frac{\partial\dot{\gamma}}{\partial{z}}V_i,
\end{equation}

\noindent with $f^k=0.57$ and $f^g=2.28$ based on controlled-velocity results for a wide range of inertial numbers ($0.05<I<0.35$). The two terms depend on several local flow properties, including $\partial{p}/\partial{z}$, $\partial{\dot\gamma}/\partial{z}$, $p$, and $\dot\gamma$, but the relative magnitude of the two terms can vary significantly for different flow geometries. Hence, we now show that (\ref{eq:scaling_R2}) remains valid in other geometries, including confined and free surface flows, while keeping $R=2$.

\begin{figure}
  \centerline{\includegraphics[width=0.9\linewidth]{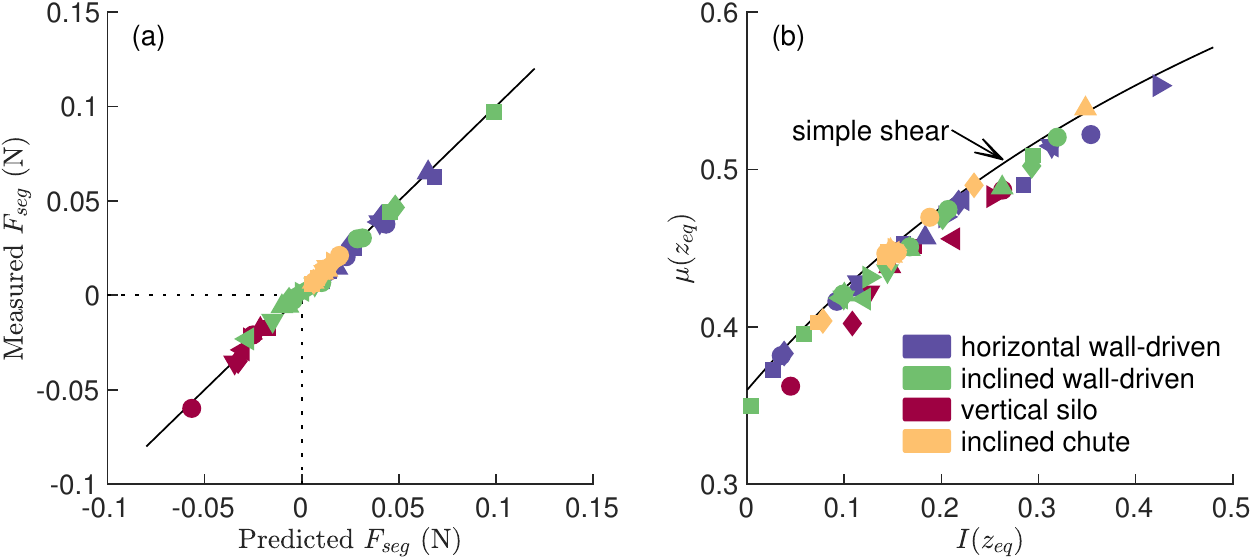}}% Images in 100% size
  \caption{
  (a) Comparison of predicted $F_{seg}$ (\ref{eq:scaling_R2}) and measured $F_{seg}$ in four flow geometries, each with widely varied system parameters (see \S\ref{sec:methods_para} for the range of parameters). The line corresponds to a perfect match between predicted and measured values. (b) Measured $\mu(I)$ from simulations at $z=z_{eq}$ compared to reference curve from simple shear simulations (see figure~\ref{fig:g0} caption). 
  }
\label{fig:unified}
\end{figure}

Following the naming convention in figure~\ref{fig:setup}, we measure $F_{seg}$ in vertical silo (confined, $g_z=0$), horizontal wall-driven (confined, $g_x=0$), inclined wall-driven (confined, $g_x\neq0$, $g_z\neq0$), and inclined chute flows (free surface), each with broadly varied system parameters covering a wide range of inertial numbers, shear rate gradients (curvatures), and pressure gradients (for $g_z\neq0$) for $R=2$. Results from $52$ simulations across all flow geometries are reported in figure~\ref{fig:unified}(a), in which the values for $F_{seg}$ measured in the simulations are plotted against predictions of (\ref{eq:scaling_R2}). The agreement between the measured values and the predictions is remarkable, especially given the broad range of flow geometries and flow conditions that are considered. The broad range of flow conditions is further amplified in figure~\ref{fig:unified}(b), where local rheological data for the cases are presented, ranging from quasistatic ($I<0.1$) to inertial flow regimes.

Three key points of figure~\ref{fig:unified} merit elaboration before the results are discussed in greater detail. First, the scaling (\ref{eq:scaling_R2}) used to predict $F_{seg}$ has only two independently determined parameters ($f^k$ and $f^g$), yet its predictions are accurate across flow geometries with widely varying boundary conditions and forcing. Second, vertical silo results of $F_{seg}$ in figure~\ref{fig:unified}(a) are always negative due to the negative curvatures of the velocity profile (figure~\ref{fig:setup}e), whereas horizontal wall-driven results of $F_{seg} $are positive due to the added effect of positive curvatures (figure~\ref{fig:setup}g) and gravity-induced contributions. For inclined wall-driven flows, $F_{seg}$ is either positive or negative due to the net effect of the flow velocity curvature (depending on $\theta$) and gravity. Inclined chute flow results in figure~\ref{fig:unified}(a) are always positive and span a much narrower range; in fact, as illustrated below, $F^k_{seg}$ is nearly negligible in free surface flows compared to $F^g_{seg}$. Third, rheological data in figure~\ref{fig:unified}(b) for different flow geometries deviate slightly from the simple shear rheology, yet this deviation does not affect the accurate prediction of $F_{seg}$ in figure~\ref{fig:unified}(a). Specifically, compared to the reference curve for simple shear, the confined flows without a controlled-velocity profile show systematic deviations that can be attributed to nonlocal effects near the edge of a localized shear layer \citep{kamrin_nonlocal_2012}. In contrast, inclined chute flows and controlled-velocity flows (see insets in figures~\ref{fig:g0} and \ref{fig:g}) closely follow the $\mu(I)$ curve for simple shear. Nevertheless, the prediction made by (\ref{eq:scaling_R2}) remains accurate despite these variations in rheology, which stands in contrast to shear stress gradient based scaling (\ref{eq:scaling1}) that is $\mu$ dependent \citep{guillard_scaling_2016}.

\begin{figure}
  \centerline{\includegraphics[width=0.9\linewidth]{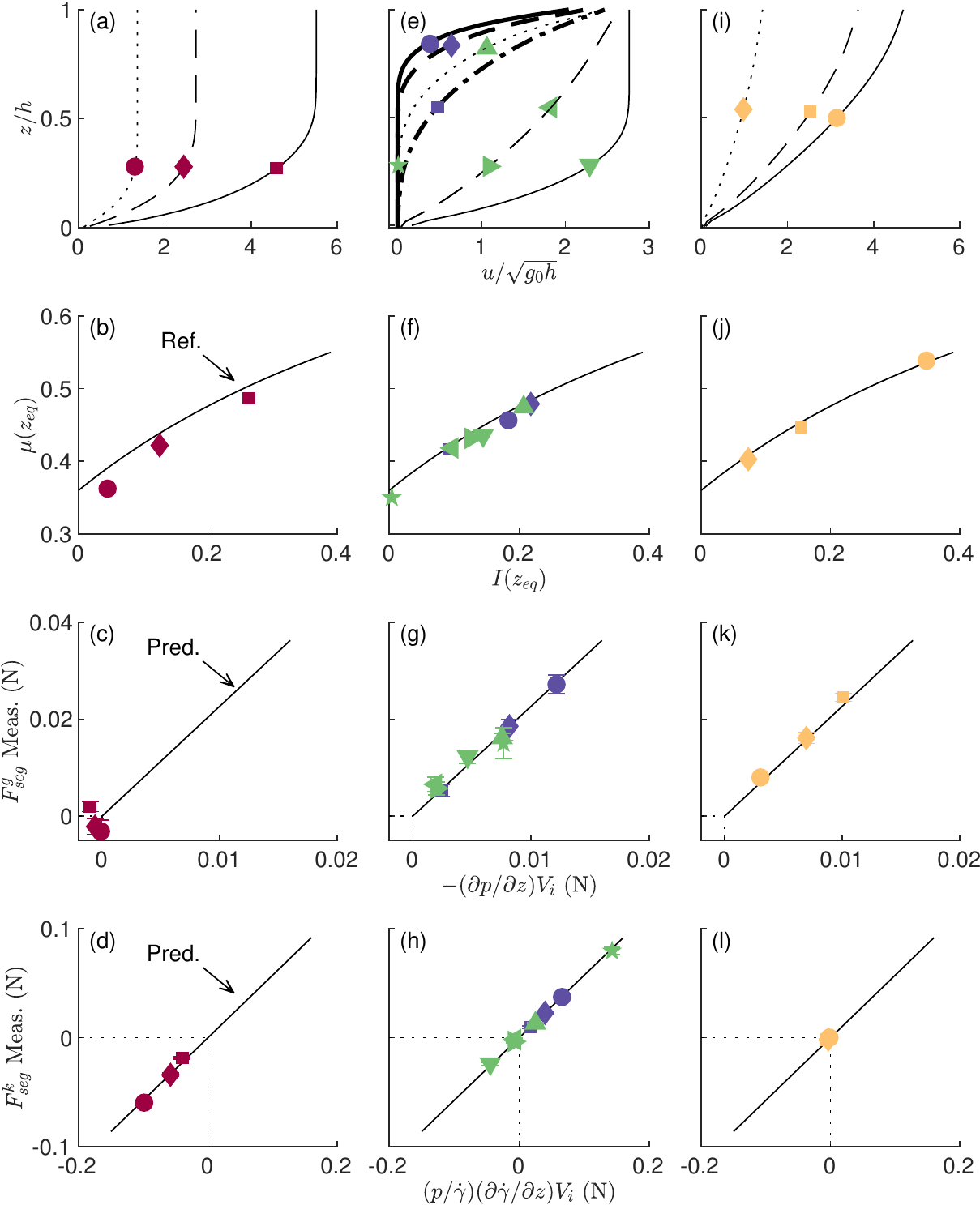}}
  \caption{
  Representative results from vertical silo flows (left), horizontal and inclined wall-driven flows (middle), and inclined chute flows (right); color schemes match those in figure~\ref{fig:unified}. Curves in the second row are reference $\mu(I)$ curves based on simple shear results. The predicted lines in the last two rows have slopes of $2.28$ and $0.57$, respectively, based on (\ref{eq:scaling_R2}).
  }
\label{fig:wall_chute}
\end{figure}

To better illustrate how gravity and flow kinematics contribute to the net segregation force in different flow geometries, we select representative cases for each geometry and present their velocity profiles, rheological data, and the two components of $F_{seg}$ in figure~\ref{fig:wall_chute}. A length scale $h$ and a velocity scale $\sqrt{g_0h}$ are consistently used for normalization, where $g_0=9.81$~\si{m/s^2} is fixed even though $g$ is case specific.

\emph{Vertical silo flows.} Figure~\ref{fig:wall_chute}(a) shows velocity profiles (curves) and steady-state locations of the intruder (symbols) for three (out of a total of eight) representative vertical silo simulations (with varying $u_0$ but the same $P_0$ and $g_x$). In vertical silo flows, shear is localized near the bottom wall with a concave-up velocity profile (negative curvature) connected to a plug-flow zone. The intruder is placed in the shear zone ($z_{eq}/h\approx0.25$). Figure~\ref{fig:wall_chute}(b) shows rheological measurements around the intruder (symbols), which deviate from the simple shear reference curve (solid line) due to nonlocal effects in localized shear. The small deviation of the rheological data in this flow geometry is robust (e.g., increasing particle stiffness by a factor of ten does not change the result significantly); nevertheless, the scaling of $F_{seg}$ is unaffected by this variation in rheology.

Figures~\ref{fig:wall_chute}(c) and (d) show scaling of gravity- and kinematics-components of $F_{seg}$ in the vertical silo flow simulations. To obtain \emph{measured} $F^g_{seg}$ and $F^k_{seg}$ (vertical axes), we subtract $0.57(p/\dot{\gamma})(\partial\dot{\gamma}/\partial{z})V_i$ and $-2.28(\partial{p}/\partial{z})V_i$ from the total measured $F_{seg}$, respectively, using local flow properties. The measured $F^g_{seg}$ and $F^k_{seg}$ are then compared to force scales $-(\partial{p}/\partial{z})V_i$ and $(p/\dot{\gamma})(\partial\dot{\gamma}/\partial{z})V_i$, showing excellent agreement with the predicted solid lines in figures~\ref{fig:wall_chute}(c) and (d). More specifically, data in figure~\ref{fig:wall_chute}(c) cluster around $(0,0)$ because $g_z=0$, whereas data in figure~\ref{fig:wall_chute}(d) spread along the negative portion of the predicted line because of differing negative curvatures. Therefore, the net effect of the total $F_{seg}$ for vertical silo flows is negative, meaning that an intruder tends to be pushed toward the immobile wall (high shear rate regions), agreeing with previous \emph{dense} silo flow simulations \citep{fan_phase_2011}.
% The agreement of the force scaling in figure~\ref{fig:wall_chute}(g) is rather significant given that vertical silo flows do not follow exactly the same $\mu(I)$ curve as controlled-velocity flows from which the scaling (\ref{eq:fcurv}) is established (figure~\ref{fig:wall_chute}b); this highlights the relevance of our kinematics-based approach. It is also interesting to note that the ranges of both axes of figure~\ref{fig:wall_chute}(c) are one order of magnitude greater than those of figure~\ref{fig:g0}(b), where the scaling is derived, because confined flows exhibit much stronger curvatures due to shear localization. Therefore, our scaling is likely scale invariant.

\emph{Horizontal wall-driven flows.} Thick curves in figure~\ref{fig:wall_chute}(e) are velocity profiles of selected horizontal wall-driven flows (three of $16$ simulations for this condition) that are concave down (positive curvature) in the shear zone close to the top moving wall. The selected cases differ only in $g_z$ but have the same $u_0$ and $P_0$; the intruder fluctuates either around $z_{eq}/h\approx0.75$ or $z_{eq}/h\approx0.5$ (blue symbols). Figure~\ref{fig:wall_chute}(f) shows that the rheological measurements for wall-driven flows deviate only slightly from the simple shear reference curve. This, again, does not affect the scaling of the two components of $F_{seg}$, which follow the predictions in both figures~\ref{fig:wall_chute}(g) and (h). Results of $F^k_{seg}$ for these positive-curvature velocity profiles fall along the positive portion of the predicted line in figure~\ref{fig:wall_chute}(g), adding to $F^g_{seg}$, which is also positive. This indicates that the intruder tends to be pushed upward by the net $F_{seg}$ in wall-driven flows; imbalances between $F_{seg}$ and the intruder weight ($m_ig_z$) determine whether the intruder rises or sinks \citep{jing_rising_2020}.

\emph{Inclined wall-driven flows.} Thin curves in figure~\ref{fig:wall_chute}(e) are velocity profiles of inclined wall-driven flows that transition from concave down (dotted) to concave up (dashed and solid) for $\theta=\{10,30,45\}^\circ$. The rheology (figure~\ref{fig:wall_chute}f) and $F^g_{seg}$ (green symbols in figure~\ref{fig:wall_chute}g) results are similar to normal wall-driven flows ($\theta=0$), but $F^k_{seg}$ (figure~\ref{fig:wall_chute}h) for inclined wall-driven flows varies from negative to positive due to the changed sign of $\partial\dot\gamma/\partial{z}$. Results match predictions (figures~\ref{fig:wall_chute}g and h) even when intruders are repositioned at different locations from $z_{eq}/h\approx0.25$ to $z_{eq}/h\approx0.75$, resulting in a wide range of local inertial numbers including a quasistatic case near the bottom of the flow (star symbol on the thin dotted curve in figure~\ref{fig:wall_chute}e). 

\emph{Inclined chute flows.} Finally, results of selected free-surface, inclined chute flows with $\theta=\{22,24,28\}^\circ$ and corresponding $g=\{5,9.81,15\}$~\si{m/s^2} are presented in figures~\ref{fig:wall_chute}(i--l), covering a wide range of inertial numbers and pressure gradients. The Bagnold-type velocity profiles are always concave up (figure~\ref{fig:wall_chute}i), and local rheological measurements match the simple shear reference curve (figure~\ref{fig:wall_chute}j). Interestingly, as shown in figure~\ref{fig:wall_chute}(l), the rescaled local curvatures are always nearly zero even though the shear rate varies extensively (characterized by the span of $I$ in figure~\ref{fig:wall_chute}j). As such, $F^k_{seg}$ is negligible compared to $F^g_{seg}$ (figure~\ref{fig:wall_chute}k). This finding matches our recent results \citep{jing_rising_2020} showing that segregation forces in inclined chute flows scale primarily with pressure gradients but not shear rate gradients.

\subsection{Variation of segregation force with $R$}

\subsubsection{Scaling based on controlled-velocity results}

\begin{figure}
  \centerline{\includegraphics[width=0.9\linewidth]{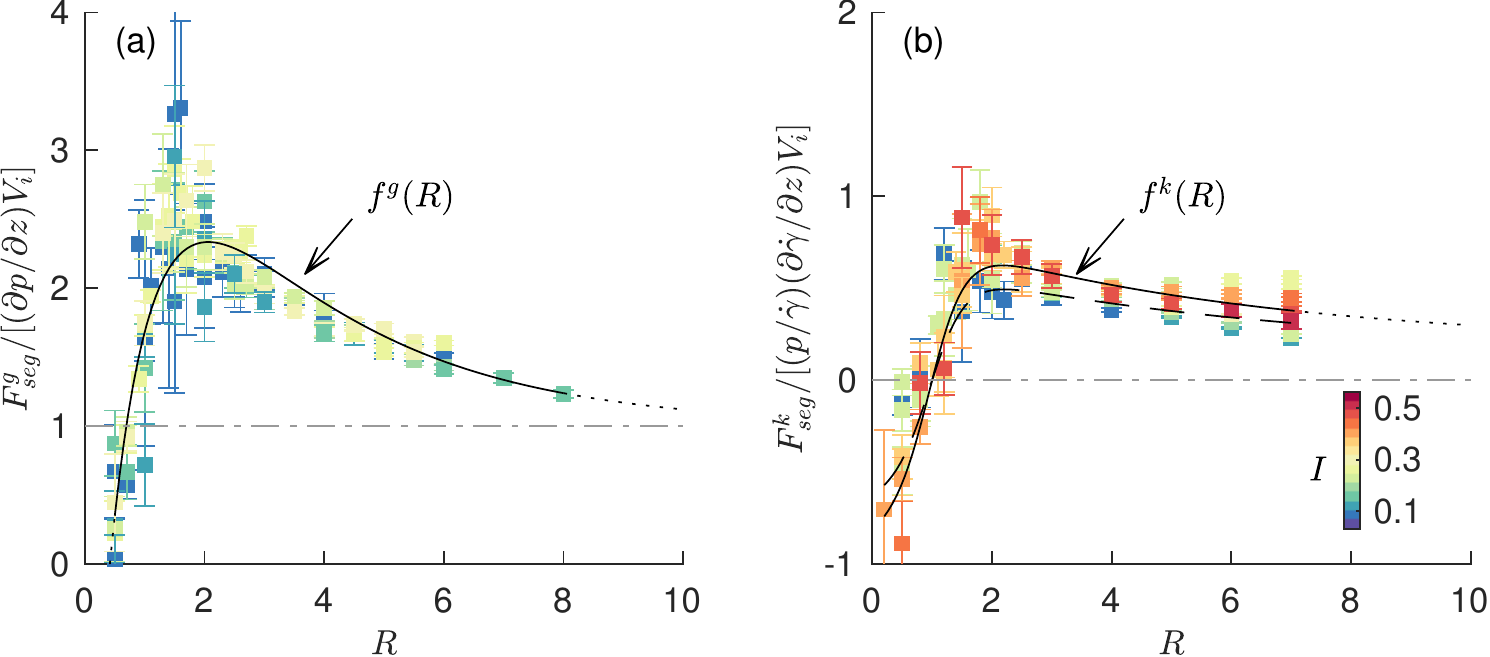}}
  \caption{
  Pre-factors of (a) gravity- and (b) kinematics-related components of $F_{seg}$ vs.\ $R$ for controlled-velocity flows. Data in (a) are from \citet{jing_rising_2020} where constant-shear-rate flows with gravity ($C_0=0$, $g=9.81$~\si{m/s^2}) are used,
  % along with an additional set of simulations for $I\approx0.56$ (red),
  while simulations in (b) are constant-curvature, no-gravity flows ($C_0h/\dot\gamma_0=\pm2$, $g=0$). In both configurations, $\dot\gamma_0$ and $P_0$ are varied to generate a range of local inertial numbers (color bar). The solid curves in (a) and (b) are semi-empirical fits described by (\ref{eq:fg}) and (\ref{eq:fk}), respectively, and the dotted curves indicate unexplored ranges of $R$. The dashed curve in (b) is a fit based only on data with $P_0\geqslant1000$~\si{Pa} (see \S\ref{sec:fR_valid}). Dash-dotted horizontal lines in (a) and (b) indicate reference values for $f^g(R)$ and $f^k(R)$, respectively, when $R=1$ (monodisperse flow).
  }
\label{fig:fR}
\end{figure}

To this point, the fitting constants $f^g$ and $f^k$ in scaling law (\ref{eq:scaling_R2}) are specific to $R=2$. However, we previously showed \citep{jing_rising_2020} in constant-shear-rate flows ($F^k_{seg}=0$) and free surface flows ($F^k_{seg}\approx0$) that the scaling factor for $F^g_{seg}$ is a function of $R$; see also (\ref{eq:scaling2}). To illustrate this, our data for $f^g:=F^g_{seg}/[-(\partial{p}/\partial{z})V_i]$ \citep{jing_rising_2020} are reproduced in figure~\ref{fig:fR}(a) for $0.05\leqslant I \leqslant 0.24$.
% together with an additional set of simulations for $I\approx0.56$ ($C_0=0$, $g=9.81$~\si{m/s^2}).
With this background, we now explore $F^k_{seg}$ for $0.2\leqslant R \leqslant 7$ using controlled-velocity, constant-curvature flows, where gravity is turned off ($F^g_{seg}=0$), $P_0$ and $\dot\gamma_0$ are varied for a wide range of $I$ ($0.03\leqslant I \leqslant 0.43$), and $C_0h/\dot\gamma_0=\pm2$ is fixed to control the local curvature. In figure~\ref{fig:fR}(b), results for $f^k:=F^k_{seg}/[(p/\dot\gamma)(\partial\dot\gamma/\partial{z})V_i]$ plotted versus $R$ collapse onto a master curve, despite widely varying local inertial numbers (indicated by colors). Note that the $f^k(R)$ data diverges somewhat for $R\gtrsim4$, but with no systematic dependence on $I$; this diverging behavior is discussed in \S\ref{sec:fR_valid}. In general, the $f^k(R)$ curve is negative for $R<1$, increases toward a peak at $R\approx2$, and declines as $R$ is further increased; its shape is similar to its gravity-related counterpart $f^g(R)$ in figure~\ref{fig:fR}(a).

The forms of $f^g(R)$ and $f^k(R)$ have interesting implications in relation to the underlying physics of gravity- and kinematics-induced segregation. As we previously noted \citep{jing_rising_2020}, the gravity-induced segregation force approaches Archimedes' buoyancy force in the continuum limit (i.e., $f^g(\infty)\rightarrow1$); this force appears to be enhanced by finite-size effects and the discrete nature of particle contacts for $1\lesssim R \lesssim 10$ but weakened by kinetics-driven percolation for $R\lesssim1$, giving rise to the peak at $R\approx2$.

The kinematics-induced segregation force might follow a similar geometric argument but with some unique characteristics. A key observation is that $f^k(R)$ changes sign at $R\approx1$. Indeed, no net segregation force (perpendicular to shear) is expected for $R=1$, which is simply a monodisperse flow. The physical implications of the negative and positive parts of $f^k(R)$ for $R<1$ and $R>1$, respectively, are discussed below.

\emph{Collision dominant for small intruders.} For $R<1$, $f^k(R)$ is negative, indicating that a small intruder is pushed from higher shear rate regions toward lower shear rate regions; this is reminiscent of a collisional argument in dilute granular flows whereby particles migrate from higher to lower granular temperatures \citep{jenkins_segregation_2002,trujillo_segregation_2003,fan_phase_2011}. Although the flows described here are dense ($I\lesssim0.5$), contact forces acting on a \emph{small} intruder are primarily collisional (as opposed to enduring) as a small intruder tends to percolate through voids \citep{silbert_rheology_2007,jing_micromechanical_2017}. As a result, small intruders experience larger contact forces from the ``hotter'' (higher shear rate and hence higher granular temperature) side and are thereby pushed toward the ``cooler'' side, corresponding to $f^k(R)<0$ for $R<1$.

\emph{Friction dominant for large intruders.} For $R>1$, intruder particles tend to undergo more enduring frictional contacts rather than collisional contacts \citep{fan_phase_2011,jing_micromechanical_2017}. Enduring contact forces are expected to be enhanced by a lower shear rate due to the increase in force correlation length, or cluster size \citep{lois_emergence_2006,fan_phase_2011}, as well as the increase in contact duration \citep{silbert_rheology_2007}. As a result, \emph{large} intruders experience larger contact forces from lower shear rate regions and are thereby pushed toward higher shear rate regions, which corresponds to $f^k(R)>0$ for $R>1$. The peak of $f^k(R)$ at $R\approx2$ indicates that the frictional effect is enhanced due to finite-size effects similar to those for $f^g(R)$ \citep{jing_rising_2020}, which tend to diminish as $R\rightarrow\infty$ (toward the continuum limit).

\emph{$R\rightarrow\infty$ limit for $f^k(R)$.} When $R\rightarrow\infty$, the flow of bed particles surrounding an intruder can be viewed as a continuum (a complex fluid), and it is intriguing to consider if the segregation force has a parallel in Newtonian fluid flows. However, unlike the Archimedian buoyancy limit for the gravity-induced segregation force, i.e., $f^g(\infty)\rightarrow1$ (see figure~\ref{fig:fR}a), the behavior of $f^k(R)$ for $R\rightarrow\infty$ is less straightforward (figure~\ref{fig:fR}b). The force (perpendicular to shear and toward higher shear rates) induced by the curvature of a flow velocity profile is reminiscent of the ``shear gradient lift force,'' $F_\textrm{SG}$, for inertial focusing in microfluidics \citep{martel_inertial_2014}, which is also known as the \emph{tubular pinch} effect \citep{segre_radial_1961,matas_inertial_2004}. Despite the complex dependence of $F_\textrm{SG}$ on the flow Reynolds number $\textrm{Re}$ and the particle distance to the channel wall $z/D_h$, where $D_h$ is the hydraulic diameter of the channel, a scaling law for $F_\textrm{SG}$ has been derived for small particle Reynolds numbers \citep{asmolov_inertial_1999} and empirically determined in microfluidic systems \citep{di_carlo_particle_2009} as $F_\textrm{SG}=C_\textrm{SG}\rho_fU_m^2a^3/D_h$, where $\rho_f$ is the fluid density, $U_m$ is the maximum channel flow velocity, $a$ is the particle radius (i.e., $a=d_i/2$ here), and the lift coefficient $C_\textrm{SG}$ is a dimensionless function of $\textrm{Re}$, $z/D_h$, and $a/D_h$. Note that for channel flows of a Newtonian fluid, the velocity profile is typically parabolic (similar to our velocity profile (\ref{eq:uz})) with $U_m$ appearing at the centerline and has a constant shear rate gradient $\partial{\dot\gamma}/\partial{z} \propto U_m/D_h^2$. For a shear rate scale $U_m/D_h$ and a (hydrodynamic) pressure scale $\rho_fU_m^2$, we have $F^k_{seg}\propto(p/\dot\gamma)(\partial{\dot\gamma}/\partial{z})V_i^3\propto\rho_fU_m^2a^3/D_h$, consistent with the scaling of $F_\textrm{SG}$. Therefore, it seems plausible to connect our kinematics-induced segregation force (in the large $R$ limit) with the shear gradient lift force that is partly responsible for cross-streamline migration of particles in confined fluid flows from lower to higher shear rate regions. Of course, further investigation is warranted to shed more light on this connection as well as to consider other inertial lift forces as a possible continuum limit for $F^k_{seg}$, such as those induced by wall effects \citep{martel_inertial_2014,ekanayake_lift_2020}, slip velocity \citep{asmolov_inertial_1999,ekanayake_lift_2020}, and the Saffman lift effect \citep{saffman_lift_1965,van_der_vaart_segregation_2018}. Nevertheless, based on the understanding and scaling arguments above, we assume that $f^k(R)$ approaches a positive finite value (denoted as $f^k_\infty$) as $R\rightarrow\infty$ and use this assumption to constrain our empirical fit in the subsequent analysis.

 % Although a sphere in a viscous shear flow is known to experience a ``lift'' force in the direction perpendicular to shear, the lift force is typically related to the shear rate \citep{saffman_lift_1965}, slip velocity (velocity lag between the sphere and the flow) \citep{saffman_lift_1965,ekanayake_lift_2020}, and, in the presence of shear rate gradients, the distance of the sphere to a wall \citep{asmolov_inertial_1999}. However, none of these factors are of direct relevance to our scaling. For example, our results show that $F^k_{bed}$ is zero and independent of the shear rate if the velocity profile is linear (see figure~\ref{fig:g0}), and our intruders are typically placed far from a rigid wall. Therefore, we do not have a clear understanding or argument for the appropriate value for $f^k(\infty)$, except that it appears to be non-zero due to the asymmetry in the associated shear rate profiles.

It is convenient to fit the collapsed data in figure~\ref{fig:fR} to functions, but it is unclear how to derive such functions analytically because they would need to bridge two unrelated physical phenomena: brief intermittent collisions at small $R$ and enduring multiple contacts at large $R$. In our previous work \citep{jing_rising_2020}, we use a double-exponential function to fit the data in figure~\ref{fig:fR}(a) based on the understanding that the two geometric effects explained above (percolation for $R\lesssim1$ and enhanced frictional contacts for $R\gtrsim1$) tend to saturate as $R$ increases, and that exponential decays are not uncommon in segregation phenomena \citep{savage_particle_1988,khola_correlations_2016,schlick_modeling_2015}. The semi-empirical fit for $f^g(R)$ is

\begin{equation}\label{eq:fg}
	f^g(R) = \Big[1-c^g_1\exp\Big({-\frac{R}{R^g_1}}\Big)\Big]\Big[1+c^g_2\exp\Big({-\frac{R}{R^g_2}}\Big)\Big],
\end{equation}

\noindent where $R^g_1=0.92$, $R^g_2=2.94$, $c^g_1=1.43$, and $c^g_2=3.55$ are fitting parameters \citep{jing_rising_2020}. A detailed explanation for the expression (\ref{eq:fg}), along with evidence for an exponential decay of an enhanced contact number density around the intruder as $R$ increases, is provided by \citet{jing_rising_2020}. Briefly, the first term accounts for necessary percolation at small $R$ and the second term accounts for enhanced enduring contacts acting on the intruder at large $R$.

Here we fit the data for $f^k(R)$ in figure~\ref{fig:fR}(b) to a functional form similar to (\ref{eq:fg}):

\begin{equation}\label{eq:fk}
	f^k(R) = f^k_\infty\Big[\tanh\Big(\frac{R-1}{R^k_1}\Big)\Big]\Big[1+c^k_2\exp\Big({-\frac{R}{R^k_2}}\Big)\Big],
\end{equation}

\noindent where $f^k_\infty=0.19$, $R^k_1=0.59$, $R^k_2=5.48$, and $c^k_2=3.63$ are fitting parameters. Compared to (\ref{eq:fg}), a non-unity pre-factor $f^k_\infty$ is used to match the assumption that $f^k(\infty)\rightarrow{f}^k_\infty$, and the first exponential function is replaced by a hyperbolic tangent function (i.e., $\tanh[(R-1)/R^k_1]$, which passes through zero at $R=1$ and approaches $1$ and $-1$ for $R\gtrsim2$ and $R\rightarrow0$, respectively) to better match the data for $0<R\lesssim2$ in figure~\ref{fig:fR}(b). Although the latter modification is phenomenological, it is interesting to note that the fitted ``shape factor'' $R^k_1$ for the $\tanh$ function in (\ref{eq:fk}) is similar to $R^g_1$ in the first exponential function of (\ref{eq:fg}) in that the values are similar in magnitude and less than one. The similarity comes about because both indicate that the geometric effects related to small intruders (percolation and collision dominant for $R<1$) decay with a ``characteristic size ratio'' of order one. Likewise, the second exponential function of (\ref{eq:fg}) and (\ref{eq:fk}) each decay with a ``characteristic size ratio'' ($R^g_2$ and $R^k_2$, respectively) that has a physically relevant value, considering the observation that the contact number density around an intruder increases sharply with $R$ for $R\lesssim3$ and saturates for $R\gtrsim5$; see \citet{jing_rising_2020}.

Despite the physical intuition associated with our semi-empirical models (\ref{eq:fg}) and (\ref{eq:fk}), one might choose other forms to fit the data in figure~\ref{fig:fR}.
% However, other functions may require more fitting parameters (four for each model here) for an equivalent fit and limiting values.
% , each with a physically appropriate value, to describe the complex non-monotonic trend of the data (e.g., with a maximum at $R\approx2$) with well constrained limiting behaviors at large and small $R$. 
Ideally, a theoretically grounded approach for determining $f^g(R)$ and $f^k(R)$ may be possible, perhaps one that extends kinetic theories for segregation \citep{jenkins_segregation_2002,trujillo_segregation_2003,duan_segregation_2020} into the dense limit.

\subsubsection{Validation of $R$-scaling in other flow geometries}\label{sec:fR_valid}

\begin{figure}
  \centerline{\includegraphics[width=0.5\linewidth]{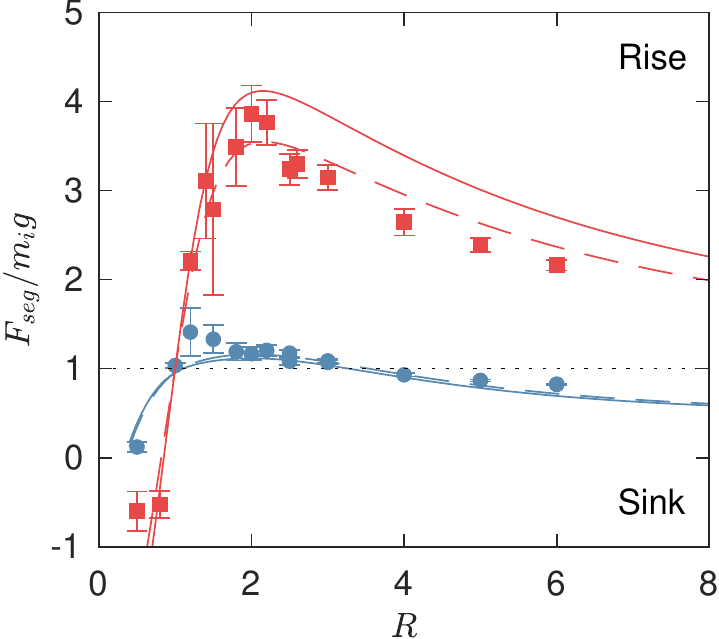}}
  \caption{ 
  Validation of predicted $F_{seg}/m_ig$ vs.\ $R$ in inclined chute flow (blue circles) ($h/d\approx40$, $\theta=24^\circ$, and $g=9.81$~\si{m/s^2}) and wall-driven flow (red squares) ($h/d\approx40$, $P_0=1500$~\si{Pa}, $\dot\gamma_0=20$~\si{s^{-1}}, and $g=3$~\si{m/s^2}) with varying $R$. Solid curves are predictions based on the scaling law (\ref{eq:scaling}) and the empirical fitting functions (\ref{eq:fg}) and (\ref{eq:fk}). Dashed curves are predictions based on a refitted $f^k(R)$ curve considering only a subset of data with $P_0\geqslant1000$~\si{Pa} (see figure~\ref{fig:fR} and text). Model inputs for predictions are \emph{local} flow properties at the height of the intruder $z_{eq}\approx0.5h$ taken from corresponding simulations. Areas above and below the dotted horizontal line at $F_{seg}/m_ig=1$ indicate where the intruder tends to rise and sink, respectively.
  % The shaded (unshaded) area indicates where the intruder tends to sink (rise).
  }
\label{fig:rise_sink}
\end{figure}

Although the unified scaling (\ref{eq:scaling}) is based on a wide range of flow conditions in figure~\ref{fig:setup}(a-c), the dependence of the semi-empirical relations (\ref{eq:fg}) and (\ref{eq:fk}) on $R$ shown in figure~\ref{fig:fR} are exclusively based on controlled-velocity confined flow conditions (figure~\ref{fig:setup}a), where parameters for $f^g(R)$ and $f^k(R)$ are calibrated separately by eliminating velocity profile curvatures and gravity, respectively. To validate the $R$-scaling for flow geometries where gravity and flow curvatures coexist, we focus on an inclined chute flow and a horizontal wall-driven confined flow, varying $R$ in each case and comparing the predicted $F_{seg}$ with simulation results (figure~\ref{fig:rise_sink}).

First, consider results for varying $R$ in an inclined chute flow ($\theta=24^\circ$, $g=9.81$~\si{m/s^2}) in figure~\ref{fig:rise_sink}, where local flow properties at $z=z_{eq}$ are $-\partial{p}/\partial{z}=1.3\times10^4$~\si{Pa/m} and $(p/\dot\gamma)(\partial\dot\gamma/\partial{z})=-0.6\times10^4$~\si{Pa/m}. The resulting kinematic term is nearly ten times smaller than the gravity term, noting that the pre-factor $f^k$ is about one-fourth of $f^g$ (figure~\ref{fig:fR}). Calculating the appropriate pre-factors using (\ref{eq:fg}) and (\ref{eq:fk}) based on $R$ and substituting them into (\ref{eq:scaling}) yields a prediction (blue solid curve in figure~\ref{fig:rise_sink}) that agrees well with simulation results (blue circles). Note that we report $F_{seg}/m_ig_z$ in figure~\ref{fig:rise_sink} to indicate whether an intruder tends to rise or sink; if $F_{seg}/m_ig_z>1$, the segregation force is greater than the intruder weight and thereby pushes the intruder upward, whereas if $F_{seg}/m_ig_z<1$, the segregation force is insufficient to support the intruder weight and the intruder sinks. For the inclined chute flow, a large intruder with $1<R<4$ tends to rise, while a small intruder ($R<1$) sinks, agreeing with the tendency of ``normal'' segregation. However, for $R>4$ the intruder tends to sink due to its large weight, known as ``reverse'' segregation \citep{felix_evidence_2004}.

For the second case of a wall-driven flow ($P_0=1500$~\si{Pa}, $\dot\gamma_0=20$~\si{s^{-1}}, $g=3$~\si{m/s^2}) with local flow properties $-\partial{p}/\partial{z}=0.4\times10^4$~\si{Pa/m} and $(p/\dot\gamma)(\partial\dot\gamma/\partial{z})=3.3\times10^4$~\si{Pa/m}, the kinematics contribution is approximately two times larger than its gravity counterpart, and, importantly, both terms are positive. This leads to a much larger predicted segregation force (red solid curve), which agrees well with simulation results (red squares) for $R\lesssim2$. However, the agreement beyond $R>2$ is not as good. Recalling that the $f^k(R)$ data in figure~\ref{fig:fR}(b) show growing scatter for $R\gtrsim4$, the mismatch may be attributed to additional mechanisms not considered in our scaling, especially for this range of $R$. To understand this, we have examined the effects of the inertial number (which considers case-specific $\dot\gamma$ and $p$), stiffness of the virtual spring that is attached to the intruder, the thickness of the flowing layer, and the domain size (particularly in the spanwise direction), but none of these explain the mismatch. One possible mechanism lies in the overburden pressure: for $P_0\geqslant1000$~\si{Pa}, $f^k(R)$ values in figure~\ref{fig:fR}(b) tend to appear near the lower edge of the scatter, especially for $R>2$, regardless of the local shear rate varying from $10$~\si{s^{-1}} to $40$~\si{s^{-1}}. This becomes evident when fitting (\ref{eq:fk}) to only the data with $P_0\geqslant1000$~\si{Pa}, corresponding to the dashed curve in figure~\ref{fig:fR}(b). On the other hand, all wall-driven flows we consider (including those in figure~\ref{fig:rise_sink}) fall into this higher overburden pressure category (which have $P_0\geqslant1500$~\si{Pa}) in order to ensure a flowing layer that is $10d$ to $30d$ thick (wall-driven flows tend to localize near the top wall). Hence, using the fit to (\ref{eq:fk}) based on the subset of the $f^k(R)$ data with $P_0\geqslant1000$~\si{Pa} (dashed curve in figure~\ref{fig:fR}b) significantly improves the agreement of the predicted (red dashed) curve with wall-driven data in figure~\ref{fig:rise_sink}. Although we do not have an explanation for this secondary overburden pressure dependence (which is not characterized by the inertial number and is reminiscent of nonlocal effects in granular flows), it appears to be only significant for very large intruders in shear-dominant cases. It might be plausible to examine if local velocity fluctuations (relative to local pressure) play a role \citep{kim_power-law_2020}, but this is beyond the scope of this work. Note that the prediction for the inclined chute flow (blue dashed curve in figure~\ref{fig:rise_sink}) is not affected by this treatment, because the kinematics contribution associated with $f^k(R)$ is negligible in free surface flows \citep{jing_rising_2020}.

% Indeed, a much larger segregation force (dashed curve) is predicted, which agrees with simulation results (open symbols) for $R\leqslant2$. However, the model overestimates the segregation force for $R>2.5$. This result is not surprising because the intruder size is comparable to the thickness of the layer of flowing particles, recalling that in this situation (figure~\ref{fig:wall_chute}e) the majority of particles near the moving upper wall move as a plug, leaving a layer only about $10d$ thick near the lower static part of the bed. This localized shear effect likely alters the segregation force. 
Nevertheless, we stress that even without the overburden pressure-specific treatment, the overall agreement in figure~\ref{fig:rise_sink} is still promising, especially given that the prediction for two very different flow geometries (wall-driven and free surface) is based purely on independent velocity-controlled flows, and the discrepancy is not larger than the inherent scatter of the data in figure~\ref{fig:fR}. More interestingly, it is clear from figure~\ref{fig:rise_sink} that $F_{seg}/m_ig_z$ in wall-driven flows is well above one for $R>1$ due to the strong \emph{positive} kinematics contribution (as predicted by our scaling law), and, as a result, the reverse segregation regime ($F_{seg}/m_ig_z<1$) is difficult to reach by varying $R$ alone; indeed, our model predicts that a crossover will occur at $R\approx13$ for this wall-driven flow. The significant difference between the rise-sink transitions indicated by $F_{seg}/m_ig_z$ in inclined chute and wall-driven flows may explain why results based on confined flow simulations alone fail to predict the sinking of very large intruders as noted by \citet{guillard_scaling_2016}, even though this behavior is observed in free surface flow experiments \citep{felix_evidence_2004}. 

% \section{Discussion}\label{sec:discussion}

% \subsection{Connection with previous work}

% \subsubsection{Shear rate gradient versus shear stress gradient}

% \subsubsection{Modified buoyancy force versus lift force}

% \subsubsection{Dense versus dilute flow regimes}

% \subsection{Future work}

% \subsubsection{Tethered versus untethered intruders}

% \subsubsection{From single intruders to bi-disperse mixtures}

\section{Conclusions}\label{sec:conclusion}

In this paper, we have used extensive DEM simulations to develop a unified description of the gravity- and kinematics-induced segregation forces on an intruder particle in dense granular flows. It is based on a fundamental scaling law (\ref{eq:scaling}) that has been validated for flows in various confined and free surface geometries. The scaling law has two additive terms, one related to gravity (buoyancy-like) and the other related to flow kinematics (specifically, the shear rate gradient); semi-empirical pre-factors for both terms depend only on the particle size ratio but not local flow properties, although the overburden pressure might cause a slight secondary effect for very large intruders. The relative significance of the two contributions vary, and this unified description of segregation forces enhances our understanding of the tendency and physical origin of segregation in different flow geometries. For free surface flows where the velocity profile typically has small shear rate gradients, the gravity-induced segregation force dominates. In fact, rising and sinking of intruders in free surface flows can be predicted by comparing the size-corrected buoyancy force alone with the intruder weight, resulting in a phase diagram determined only by the size and density ratios \citep{jing_rising_2020}. For vertical silo flows where segregation is normal to gravity, kinematics-induced segregation forces cause large (small) intruders to migrate toward (away from) the wall, consistent with previous results \citep{fan_phase_2011}. Both mechanisms are significant in wall-driven flows (where they cooperate) and inclined wall-driven flows (where they either cooperate or compete, depending on the angle of inclination).

The physical origin of the kinematics-induced segregation force is clarified by considering its shear rate gradient-based scaling and the nature of contact forces acting on an intruder: for large intruders ($R>1$), which experience enduring frictional forces, larger forces are possible for lower shear rates (more persistent), whereas for small intruders ($R<1$), which preferentially receive collisional forces, larger forces occur in higher shear rate regions (more impulsive). It is also interesting that for wall-driven flows, which are relevant to annular shear \citep{golick_mixing_2009} and fluid-driven bedload transport \citep{ferdowsi_river-bed_2017}, both segregation-driving mechanisms (gravity- and kinematics-induced) act in the same direction against gravity and, therefore, sinking of large particles due to weight effects does not occur as readily as in free surface flows \citep{felix_evidence_2004}. From a practical standpoint, our results suggest that it may be possible to design experimental or industrial devices that minimize segregation (or, equivalently, enhance particle mixing) by manipulating the velocity profile (e.g., inclined wall-driven flows with appropriate inclinations), such that the kinematics-induced contribution to the segregation force counteracts gravitational contributions.

The unified description of segregation forces presented here is based on the limit where the intruder concentration approaches zero. However, the question naturally arises as to its applicability at higher intruder concentrations. We have already demonstrated \citep{jing_rising_2020} that it compares well with free surface flow experiments \citep{felix_evidence_2004} for larger concentrations (about $10\%$) where intruder particles interact with each other infrequently and gravity-induced segregation forces dominate over kinematics-induced forces. Recent work also indicates that under certain conditions combined size and density segregation can be nearly independent of either the particle size or density ratio at small concentrations, but further increasing the concentration leads to significant changes in the segregation behavior \citep{duan_modelling_2020}. Understanding how segregation forces depend on the particle concentration is of great interest due to the complex concentration dependence of segregation fluxes \citep{gajjar_asymmetric_2014,van_der_vaart_underlying_2015,gray_particle-size_2015,jing_micromechanical_2017,jones_asymmetric_2018,duan_modelling_2020}. Extending the current model toward finite intruder particle concentrations has the potential to further elucidate the underlying physics of granular segregation as well as to enhance continuum modeling of granular segregation, which will be addressed in future work. 

Finally, the segregation force studied in this paper, as well as in \citet{guillard_scaling_2016}, \citet{van_der_vaart_segregation_2018}, and \citet{ jing_rising_2020}, is measured on intruder particles tethered to a virtual spring, which effectively prevents segregation. That is, the \emph{mean} relative velocity between the intruder and bed particles is zero in the segregation ($z$) direction once the intruder reaches an equilibrium $z$-position. If the intruder is \emph{untethered} and the net force is unbalanced, a relative velocity between the intruder and its surrounding bed particles will develop \citep{tripathi_numerical_2011,staron_rising_2018}, leading to a resistive force (often viewed as the drag force) that can be associated with the relative velocity as well as other parameters. Previous studies on the drag force during particle segregation have focused on density-bidisperse but size-monodisperse flows \citep{tripathi_numerical_2011,duan_segregation_2020} due to the lack of a general description of the segregation force. With the segregation force model developed here, it is now possible to study the drag force in more general segregation situations where particles differ in both size and density.

\section*{Acknowledgements}
We thank Yi Fan, John Hecht, and Yifei Duan for valuable discussions. This material is based upon work supported by the National Science Foundation under Grant No. CBET-1929265, and was facilitated in part by the computational resources and staff contributions provided by the Quest high performance computing facility at Northwestern University, which is jointly supported by the Office of the Provost, the Office for Research, and Northwestern University Information Technology.

\appendix

\section{Connection with a previous model}\label{appA}

Here we demonstrate how our $\partial\dot\gamma/\partial{z}$-based scaling (\ref{eq:scaling}) is connected with the previous scaling (\ref{eq:scaling1}) where $\partial\tau/\partial{z}$ is used \citep{guillard_scaling_2016} when a local rheology is assumed. For simplicity, we adopt a linear rheological law,

\begin{equation}\label{eq:muI}
	\mu(I)=\mu_s+bI,
\end{equation}

\noindent where $\mu_s$ and $b$ are constants. The linear form \citep{da_cruz_rheophysics_2005} is a good approximation of our $\mu(I)$ data, at least for $I<0.3$ (see figure~\ref{fig:unified}). 

Expressing $\partial\tau/\partial{z}$ using (\ref{eq:muI}) with $\tau=\mu p$ and $I=\dot\gamma d/\sqrt{p/\rho}$, we have

\begin{equation}\label{eq:diff}
	\frac{\partial\tau}{\partial z} = \mu\frac{\partial p}{\partial z}+p\frac{\partial\mu}{\partial z} = \mu\frac{\partial p}{\partial z}+p\frac{\partial}{\partial z}\Big(b\frac{\dot\gamma d}{\sqrt{p/\rho}}\Big).
\end{equation}

Expanding (\ref{eq:diff}) and eliminating $b$ by identifying $b=(\mu-\mu_s)/I$ yields

\begin{equation}\label{eq:dtaudz}
	\frac{\partial\tau}{\partial z} = \frac{1}{2}(\mu+\mu_s)\frac{\partial p}{\partial z} + (\mu-\mu_s)\frac{p}{\dot\gamma}\frac{\partial\dot\gamma}{\partial z}.
\end{equation}

Rearranging and substituting (\ref{eq:dtaudz}) into (\ref{eq:scaling}), we have 
\begin{equation}
	F_{seg}=-\underbrace{\Big[ f^g(R)+\frac{1}{2}f^k(R)\frac{\mu+\mu_s}{\mu-\mu_s} \Big]}_{\sim \mathcal{A}(\mu,R)} \frac{\partial p}{\partial z}V_i + \underbrace{\frac{f^k(R)}{\mu-\mu_s}}_{\sim \mathcal{B}(\mu,R)} \frac{\partial \tau}{\partial z}V_i,
\end{equation}

\noindent where $f^g(R)$ and $f^k(R)$ are pre-factors of our scaling (\ref{eq:scaling}), while the under-braced terms are rational functions of $\mu$ with shapes similar to $\mathcal{A}(\mu,R)$ and $\mathcal{B}(\mu,R)$ in (\ref{eq:scaling1}), which are described by exponential functions in \citet{guillard_scaling_2016}.

From this analysis, it is clear that the $\mu$ dependence in (\ref{eq:scaling1}) emerges as a result of the adopted $\mu(I)$ rheology that connects $\tau$ and $\dot\gamma$ via a unique function. Although our kinematics-based scaling (\ref{eq:scaling}) can be transformed into the stress-based scaling (\ref{eq:scaling1}) with certain assumptions, we note that our scaling decouples gravity- and shear-induced (or kinematics-induced) segregation forces. Consequently, each of the resulting terms requires fewer fitting parameters (insensitive to $\mu$). It also seems to be more general across flow geometries where $\mu(I)$ might not follow a unique function, which is often the case for confined flows or near the edge of flowing regions. Finally, the shear-rate-gradient-based scaling helps reveal important aspects of the underlying physics of kinematics-induced segregation in dense granular flows.

\bibliographystyle{jfm}
\bibliography{reference}

\end{document}